\documentclass[
    ,final            
  ]
  {myaipproc}

\layoutstyle{6x9}

\begin{document}

\title{On how to complete the Dynamical Generation of Quark-Level Linear Sigma Model like Theories beyond one Loop}


\author{Frieder Kleefeld \footnote{e-mail: kleefeld@cfif.ist.utl.pt, URL: http://cfif.ist.utl.pt/$\sim$kleefeld/ \newline \mbox{}$\;\;$ Present address for postal correspondence: Pfisterstr.\ 31, 90762 F\"urth, Germany}}{
  address={Collaborator of the\\ Centro de F\'{\i}sica das Interac\c{c}\~{o}es Fundamentais (CFIF), Instituto Superior T\'{e}cnico,\\
Edif\'{\i}cio Ci\^{e}ncia, Piso 3, Av. Rovisco Pais, P-1049-001 LISBOA, Portugal}
} 

\begin{abstract}
A self-consistent strategy is proposed to complete in a renormalization scheme independent way the dynamical generation of Quark-Level Linear Sigma Model like Lagrangean theories beyond one loop like the theories of strong and electroweak interactions. The present discussion refers for simplicity to scalar and pseudoscalar degrees of freedom only while disregarding yet --- without loss of generality --- vector and axial vector degrees of freedom. Moreover points the discussion to approximations underlying dimensional and implicit regularization as presently used.
\end{abstract}

\maketitle

\section{Dynamical generation of Lagrangean theories}
Throughout the construction of \mbox{Lagrangean} densities used e.g.\ in particle physics one faces at least two problems:  \mbox{1) unpredictive} Lagrangeans contain too many uncorrelated parameters (masses, couplings) which have to be fitted to experiment; \mbox{2) inherent} divergencies of logarithmic, linear, quadratic, $\ldots$ type need to be renormalized. The concept of {\em dynamical \mbox{generation}} \cite{Kleefeld:2005hd}
of Lagrangean theories addresses and solves both issues simultaneously:
\begin{itemize}
\item[1)] In the spirit of Eguchi \cite{Eguchi:1976iz} one starts out from very few fundamental 3-point interaction vertices and constructs then on the basis of these vertices by ``loop-shrinking''  \cite{Kleefeld:2005hd} the so-called effective action (and its underlying Lagrangean) containing also terms for all remaining n-point vertices between the fields making up the theory.
\item[2)] The couplings of the fundamental 3-point interaction vertices are then chosen such that linear, quadratic \cite{Kleppe:1992qq,Deshpande:1983ka}, $\ldots$ divergencies cancel \cite{Kleefeld:2005hd} while the remaining logarithmic divergencies are renormalized \cite{Collins:1984xc} by adding to the effective action counter terms which replace in the spirit of the log.-divergent gap equation of Delbourgo and Scadron (DS) \cite{Delbourgo:1993dk,Delbourgo:1998kg} the integral $\int d^4p \;(p^2-\overline{m}^{\,2})^{-2}$ at some experimentally defined renormalization scale $\overline{m}$ by some universal complex number.
\end{itemize}
\section{Renormalization of logarithmic divergencies} 
Throughout the manuscript we shall apply the renormalization procedure of DS \cite{Delbourgo:1993dk,Delbourgo:1998kg} and replace the log.-divergent Bosonic one-loop integral at some renormalization scale $\overline{m}$ (being in the case of DS approximately equal to the nonstrange constituent quark mass $\hat{m}$, i.e.\ $\overline{m}\simeq \hat{m}= m_q$) by the finite number $+\; \frac{i}{16\, \pi^2}$ by adding suitable counterterms to the effective action. Hence we shall perform in all log.-divergent integrals the replacement
\begin{equation} I_2(\overline{m}^{\,2}) \equiv \int \frac{d^4p}{(2\pi)^4} \;  \frac{1}{(p^2-\overline{m}^{\,2})^2}  \; \rightarrow \; +\; \frac{i}{16\, \pi^2} \; , \label{eqlogdivgeq1}
\end{equation} 
being known as the {\em log.-divergent gap equation} \cite{Delbourgo:1993dk,Delbourgo:1998kg,Hakioglu:1990kg}. The replacement can be understood as the analytical continuation of the following integral identity (see also Eq.\ (\ref{eqint1}) in Appendix A) to the the exponent $n=2$: 
\begin{equation} I_n(\overline{m}^{\,2}) \equiv \int \frac{d^4p}{(2\pi)^4} \; \frac{1}{(p^2-\overline{m}^{\,2})^n} \; \stackrel{n\ge 3}{=}\; (-1)^n\, \frac{i}{16\,\pi^2}\; \frac{1}{(n-1)!\; \overline{m}^{\,2n-4}} \; .
\end{equation}
To apply the renormalization merely in integrals with zero external four-momentum (``local integrals'') we recall here that it has been pointed out in the context of the so-called implicit regularization scheme \cite{Brizola:1998tn,Sampaio:2002ii} that divergent integrals with nonvanishing external four-momentum (``non-local integrals'') can be decomposed in equally divergent local integrals and less divergent non-local integrals by repeated application of the identity $1/((p-k)^2-m^2)=1/(p^2-m^2)-(k^2-2k\cdot p)/[(p^2-m^2)((p-k)^2-m^2)]$.     
After making divergent integrals ``local'' a subsequent repeated application of the identity
\begin{equation} \frac{1}{p^2-m^2} \; = \; \frac{1}{p^2-\overline{m}^{\,2}} + \frac{m^2-\overline{m}^{\,2}}{(p^2- \overline{m}^{\,2})(p^2-m^2)}
\end{equation}
to the local limit of propagators in the integrands will then isolate all quadratic and logarithmic divergencies such that they can either be renormalized by applying the log.-divergent gap equation or by cancelling quadratic divergencies at the renormalization scale $\overline{m}$. Following this prescription we obtain with the help of the integral identities listed in Appendix A e.g.:
\begin{eqnarray} I_1(m^2) & = & I_1(\overline{m}^{\,2}) + (m^2-\overline{m}^{\,2})\;\underbrace{I_2(\overline{m}^{\,2})}_{\rightarrow \frac{i}{16\,\pi^2}}  + (m^2-\overline{m}^{\,2})^2\;I_{2,1}(\overline{m}^{\,2},m^2) \nonumber \\[1mm]
 & \rightarrow &  I_1(\overline{m}^{\,2}) +\frac{i}{16\,\pi^2}\;(m^2-\overline{m}^{\,2}) \; \left( 2- \frac{m^2}{m^2-\overline{m}^{\,2}} \; \ln \frac{m^2}{\overline{m}^{\,2}} \right) \; , \label{eqid1} \\[4mm]
 I_2(m^2) & = & \underbrace{I_2(\overline{m}^{\,2})}_{\rightarrow \frac{i}{16\,\pi^2}} + 2\,(m^2-\overline{m}^{\,2})\;I_{2,1}(\overline{m}^{\,2},m^2)  + (m^2-\overline{m}^{\,2})^2\;I_{2,2}(\overline{m}^{\,2},m^2) \nonumber \\[1mm]
 & \rightarrow & \frac{i}{16\,\pi^2} \left( 1 - \,\ln \frac{m^2}{\overline{m}^{\,2}}\right) \; ,  \label{eqid2} \\[2mm]
 I_{1,1}(\overline{m}^{\,2},m^2) & = & \underbrace{I_2(\overline{m}^{\,2})}_{\rightarrow \frac{i}{16\,\pi^2}} + (m^2-\overline{m}^{\,2})\;I_{2,1}(\overline{m}^{\,2},m^2) \; \rightarrow \;  \frac{i}{16\,\pi^2} \; \left( 2- \frac{m^2}{m^2-\overline{m}^{\,2}} \; \ln \frac{m^2}{\overline{m}^{\,2}} \right)\; ,  \label{eqid3} \nonumber \\
 & & \\
 I_{1,1}(m^2_1,m^2_2) & = & \frac{1}{2} \, \Big( I_{1,1}(\overline{m}^{\,2},m^2_1)+ I_{1,1}(\overline{m}^{\,2},m^2_2) + (m^2_1+m^2_2-\overline{m}^{\,2})\;I_{1,1,1}(m^2_1,m^2_2,\overline{m}^{\,2})\Big) \nonumber \\[1mm]
 & \rightarrow & \frac{i}{32\,\pi^2} \; \, \Bigg( 4 - \frac{m^2_1}{m^2_1-\overline{m}^{\,2}} \; \ln \frac{m^2_1}{\overline{m}^{\,2}} - \frac{m^2_2}{m^2_2-\overline{m}^{\,2}} \; \ln \frac{m^2_2}{\overline{m}^{\,2}} \nonumber \\[1mm]
 & & + (m^2_1+m^2_2-\overline{m}^{\,2})\;\; \frac{\displaystyle m^2_1\,m^2_2 \; \ln \frac{m_1^2}{m^2_2} + m^2_2\,\overline{m}^{\,2} \; \ln \frac{m_2^2}{\overline{m}^{\,2}} + \overline{m}^{\,2}\,m^2_1 \; \ln \frac{\overline{m}^{\,2}}{m^2_1}}{\displaystyle (m_1^2-m^2_2)(m_2^2-\overline{m}^{\,2})(\overline{m}^{\,2}-m^2_1)} \; \Bigg) .\; \quad  \label{eqid4} 
\end{eqnarray}
Analogously, the renormalization of the equal mass sunset diagram with zero external momentum is performed in Appendix B.
\section{The Quark-Level Linear Sigma Model (QLL$\sigma$M)} 
For various reasons like e.g. the still lacking \cite{Kleefeld:2005hf,Kleefeld:2004jb,Kleefeld:2002au} evidence for the existence of gluons and new developments in mathematical physics there has developed an alternative approach to strong interactions being different from Quantum Chromodynamics (QCD) which is known as the so-called QLL$\sigma$M.\footnote{Historically it has been probably L\'{e}vy \cite{Levy:1967a} the first to add Fermions, i.e.\ nucleons, to the Linear Sigma Model (L$\sigma$M) of Schwinger, Gell-Mann and L\'{e}vy \cite{GellMann:1960np} while Cabbibo and Maiani \cite{Cabibbo:1970uc} presumably were the first to replace the nucleons by quarks. The name QLL$\sigma$M has been coined by Delbourgo and Scadron \cite{Delbourgo:1993dk,Delbourgo:1998kg} who undertook on the basis of dimensional regularization a dynamical generation of the QLL$\sigma$M just to one loop. It has been then the author of the manuscript to point out that the experimentally favoured assymptotically free phase of QLL$\sigma$M belongs to the well-acceptable class of theories being non-Hermitian, yet PT-symmetric \cite{Kleefeld:2005hd,Kleefeld:2002au,Kleefeld:2005at}\cite{Bender:1998ke}.} The spin 1/2 Fermions of the QLL$\sigma$M, i.e. the (anti)quarks, are not interacting via gluons like in QCD, yet via mesons described by Bosonic scalar, pseudoscalar, vector and axial-vector fields.
In order to outline the idea of how to perform a complete dynamical generation of QLL$\sigma$M like Lagrangeans beyond one loop order it is not necessary to consider the full-fledged Lagrangean of the U(6)$\times$U(6) QLL$\sigma$M \cite{Kleefeld:2005hd,Kleefeld:2002au,Kleefeld:2005at}, yet one can either restrict one-self to the much simpler Lagrangean of the SU(2)$\times$SU(2) QLL$\sigma$M studied thoroughly to one loop e.g. by DS  \cite{Delbourgo:1993dk,Delbourgo:1998kg} or the U(1)$\times$U(1) QLL$\sigma$M being intimately related to the (supersymmetric) Wess-Zumino model \cite{Wess:1973kz}.  The apparent similarity of the field content of the Lagrangeans of the QLL$\sigma$M and electroweak interactions allows to transfer and extend the ideas of this manuscript explained within the context of the QLL$\sigma$M in a straight forward way to the theory of the electroweak force and to combine the former and the latter to a new standard model of particle physics.

Disregarding here for simplicity vector and axial vector mesons the SU(2)$\times$SU(2) QLL$\sigma$M assuming $N_F=2\,N_c=6$ Fermions, one scalar isoscalar meson $\sigma$ and $N_\pi=3$ pions is constructed on the basis the interaction Lagrangean ${\cal L}_{\small\makebox{quark-meson}}(x)=g \, \overline{q^c_+}(x) \,(\sigma(x) + i \,\gamma_5 \, \vec{\tau}\cdot \vec{\pi}(x)) \,q_-(x)$ \cite{Delbourgo:1993dk,Delbourgo:1998kg,vanBeveren:2002mc} yielding by loop-shrinking the following leading terms in the Lagrangean of the effective action for meson-meson interactions :
\begin{equation} {\cal L}_{\small\makebox{meson-meson}} = g_{\sigma\pi\pi} \; \sigma(x) (\sigma(x)^2+\vec{\pi}(x)^2) -\frac{\lambda}{4} \, (\sigma(x)^2+\vec{\pi}(x)^2)^2 + \ldots \; . 
\end{equation}
Analogously the  U(1)$\times$U(1) QLL$\sigma$M assuming $N_F$ Fermions, one scalar isoscalar meson $\sigma$ and one isoscalar pseudoscalar meson $\eta$ is constructed
on the basis the interaction Lagrangean ${\cal L}_{\small\makebox{quark-meson}}(x)=g \, \overline{q^c_+}(x) \,(\sigma(x) + i \,\gamma_5 \, \eta(x)) \,q_-(x)$ yielding by loop-shrinking the following leading terms in the Lagrangean of the effective action for meson-meson interactions:
\begin{equation} {\cal L}_{\small\makebox{meson-meson}} = g_{\sigma\eta\eta} \; \sigma(x) (\sigma(x)^2+\eta(x)^2) -\frac{\lambda}{4} \, (\sigma(x)^2+\eta(x)^2)^2 + \ldots \; . 
\end{equation}
Although we are going to present in what follows analytical results for the SU(2)$\times$SU(2) QLL$\sigma$M only, the analogous results for the U(1)$\times$U(1) QLL$\sigma$M are easily recovered by setting $N_\pi=1$ and performing the replacements $g_{\sigma\pi\pi}\rightarrow g_{\sigma\eta\eta}$ and $\vec{\pi}^2\rightarrow \eta^2$.
\section{Dynamical generation of the SU(2)$\times$SU(2) QLL$\sigma$M} 
\begin{figure}
  \includegraphics[width=1.00\textwidth,height=0.08\textheight]{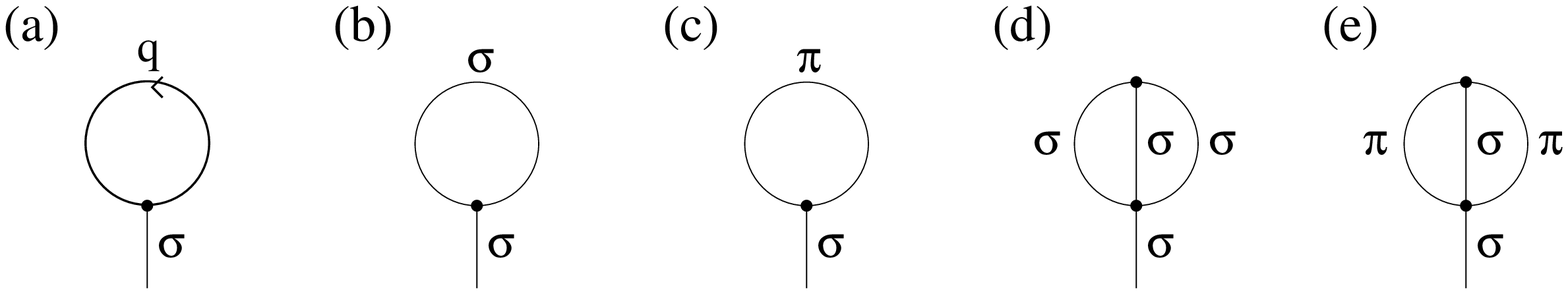}
  \caption{Tadpole sum: contributions to the $\sigma$ one-point function} \label{fig1}
\end{figure}

Following the formalism described in Ref.\ \cite{Kleefeld:2005hd} the relevant terms in the effective action of the SU(2)$\times$SU(2) QLL$\sigma$M for the $\sigma$-one-point function (see Fig.\ \ref{fig1}), for the two-point function of the quarks (see Fig.\ \ref{fig2}), of the $\sigma$  (see Fig.\ \ref{fig3}) and of the pions  (see Fig.\ \ref{fig4}) are obtained in Appendix C. Instead of determining directly the effective meson-meson-interaction couplings $g_{\sigma\pi\pi}$ and $\lambda$  completely by loop shrinking as a function of the quark-meson coupling $g$, we shall take here a different strategy and try to obtain the functional relation between the three couplings by direct elimination of quadratic divergencies in the effective action. Recalling the quadratic divergence of the sunset/sunrise diagram as discussed in Appendix B we extract the quadratically divergent part of the effective actions of Appendix C with the following result:
\begin{eqnarray} \lefteqn{S_{(1)}[\sigma] =  \int d^4x \; \sigma(x)}  \nonumber \\[1mm]
 & & \times \int \frac{d^4p}{(2\pi)^4} \frac{1}{p^2-\overline{m}^{\,2}} \; i\,\left\{ -4\,g\,N_F\,m_q +\left(1-\frac{\lambda}{4\pi^2}\right)g_{\sigma\pi\pi} (3+N_\pi)\right\} + \ldots \label{quaddiv1} \\[2mm]
 \lefteqn{S_{(2)}[\bar{q}q] = \frac{i}{2}\int d^4x \;\, \overline{q^c_+}(x)\,q_-(x)} \nonumber \\[1mm]
 & & \times \int \frac{d^4p}{(2\pi)^4} \; \frac{1}{p^2-\overline{m}^{\,2}} \; \frac{2\,g}{m^2_\sigma} \;\left\{ -4\,g\,N_F\,m_q +\left(1-\frac{\lambda}{4\pi^2}\right)g_{\sigma\pi\pi} (3+N_\pi)\right\} + \ldots  \label{quaddiv2} \\
 \lefteqn{S_{(3)}[\sigma^2] = \frac{i}{2}\int d^4x \;\, \sigma(x)^2} \nonumber \\[1mm]
 & & \times \Bigg[\int \frac{d^4p}{(2\pi)^4} \; \frac{1}{p^2-\overline{m}^{\,2}} \; \frac{6\,g_{\sigma\pi\pi}}{m^2_\sigma} \;\left\{ -4\,g\,N_F\,m_q +\left(1-\frac{\lambda}{4\pi^2}\right)g_{\sigma\pi\pi} (3+N_\pi)\right\} \nonumber  \\[1mm]
 & & \;\;\;+ \int \frac{d^4p}{(2\pi)^4} \; \frac{1}{p^2-\overline{m}^{\,2}} \;\left\{ 4\,g^2\,N_F -\left(1-\frac{\lambda}{4\pi^2}\right) \lambda\, (3+N_\pi)\right\}\Bigg] + \ldots \label{quaddiv3}  \\
 \lefteqn{S_{(4)}[\vec{\pi}^2] = \frac{i}{2}\int d^4x \;\, \vec{\pi}(x)^2} \nonumber \\[1mm]
 & & \times \Bigg[\int \frac{d^4p}{(2\pi)^4} \; \frac{1}{p^2-\overline{m}^{\,2}} \; \frac{2\,g_{\sigma\pi\pi}}{m^2_\sigma} \;\left\{ -4\,g\,N_F\,m_q +\left(1-\frac{\lambda}{4\pi^2}\right)g_{\sigma\pi\pi} (3+N_\pi)\right\} \nonumber  \\[1mm]
 & & \;\;\;+ \int \frac{d^4p}{(2\pi)^4} \; \frac{1}{p^2-\overline{m}^{\,2}} \;\left\{ 4\,g^2\,N_F -\left(1-\frac{\lambda}{4\pi^2}\right) \lambda\, (3+N_\pi)\right\}\Bigg] + \ldots  \label{quaddiv4} 
\end{eqnarray}
\noindent There are now (at least) three options to proceed:\\[2mm]
\underline{\em \bf Option 1:} {\em dynamical generation by complete elimination of quadratic divergencies}\\[2mm]
Elimination of quadratic divergencies in the tadpole-sum results in Eq.\ (\ref{eqtadpole1}), while the elimination of quadratic divergences in the remaining part of the $\sigma$- and $\pi$-self-energy yields Eq.\ (\ref{eqself1}):
\begin{eqnarray} 0 & = & -4\,g\,N_F\,m_q +\left(1-\frac{\lambda}{4\pi^2}\right)g_{\sigma\pi\pi} (3+N_\pi) \label{eqtadpole1} \\
 0 & = &  + 4\,g^2\,N_F -\left(1-\frac{\lambda}{4\pi^2}\right) \lambda\, (3+N_\pi) \label{eqself1}
\end{eqnarray}
The system of equations can be solved for $\lambda$ and $g_{\sigma\pi\pi}$ as a function of $g$ and $m_q$:
\begin{eqnarray} 0 = \frac{\lambda^2}{4\pi^2} - \lambda + \frac{4\,g^2\,N_F}{3+N_\pi} & \Rightarrow & \lambda = 2\pi^2 \left( 1 \pm \sqrt{1 - \frac{4\,g^2\,N_F}{\pi^2 (3+N_\pi)}} \;\right) \label{eqself2} \\
 \frac{4\,g\,N_F\,m_q}{g_{\sigma\pi\pi}} = \frac{4\,g^2\,N_F}{\lambda} & \Rightarrow & g_{\sigma\pi\pi} = \lambda \; \frac{m_q}{g} \label{eqlambdag1} \\
 & \Rightarrow & g_{\sigma\pi\pi} = 2\pi^2 \left( 1 \pm \sqrt{1 - \frac{4\,g^2\,N_F}{\pi^2 (3+N_\pi)}} \;\right) \; \frac{m_q}{g}
\end{eqnarray}
It's interesting to note that the identity  $g_{\sigma\pi\pi} = -\,e^{\,i\,(\alpha-\beta)}\, (m^2_\sigma-m^2_\pi)/(2\,|f_\pi|)$ of the L$\sigma$M in combination with Eq.\ (\ref{eqlambdag1}) would imply the following generalized NJL-relation:
\begin{equation} m^2_\sigma = m^2_\pi + (2\,m_q)^2  \left( -\frac{|g|\,f_\pi}{m_q} \right) \; \left( \frac{\lambda}{2\,g^2} \right) =  m^2_\pi + |2\,m_q|^2  \left( -\frac{|g|\,f_\pi}{m_q} \right) \; \left( \frac{\lambda}{2\,|g|^2} \right) \; e^{\,2\,i\,(\beta-\alpha)}\; ,  \end{equation}
which turns --- presuming the Golberger-Treiman (GT) relation on the quark-level $m_q\approx -|g|\,f_\pi$ and some eventually complex-valued quark-meson-coupling constant $g=|g|\, e^{i\alpha}$ and some complex-valued decay constant $f_\pi=|f_\pi|\,e^{i\beta}$ --- into the standard NJL-relation $m^2_\sigma \approx m^2_\pi + |2\,m_q|^2$ for $\lambda\approx 2\,|g|^2\,e^{\,2\,i\,(\alpha -\, \beta)}$ to be confronted with Eq.~(\ref{eqself2}).\\[4mm]
\clearpage

\begin{figure}
  \includegraphics[width=1.00\textwidth,height=0.16\textheight]{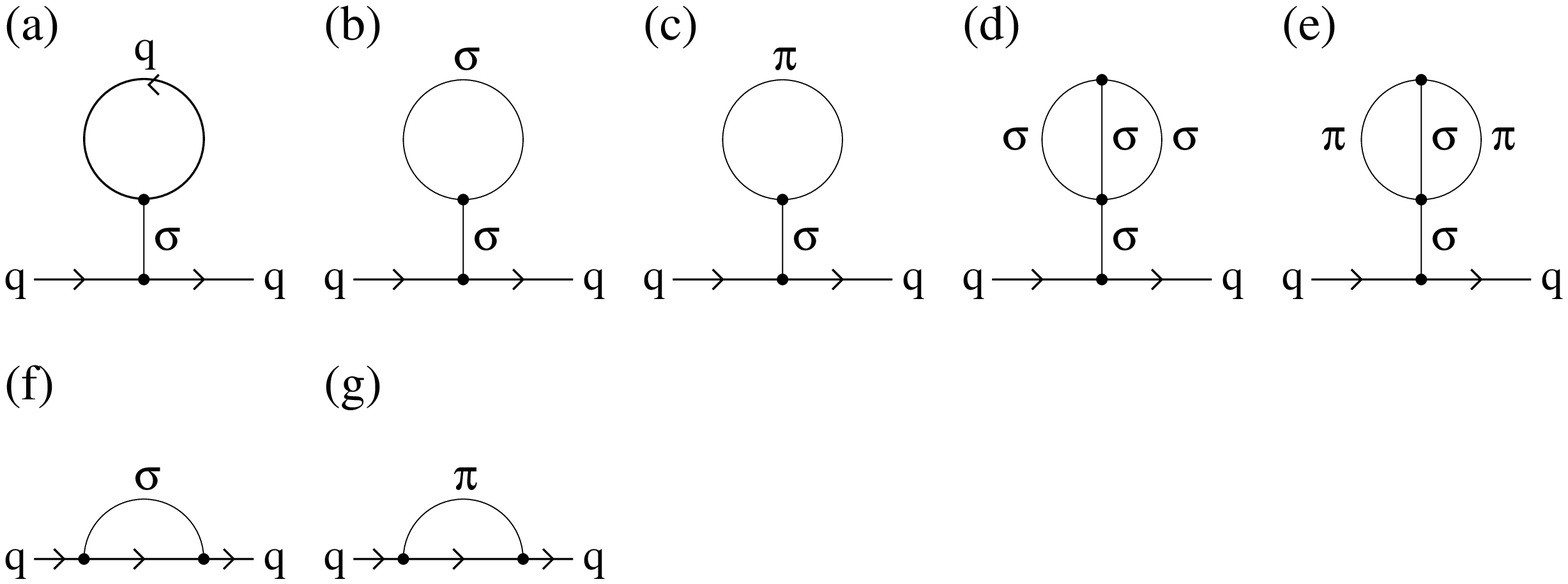}
  \caption{Quark mass: contributions to the quark self-energy} \label{fig2}
\end{figure}

\begin{figure}
  \includegraphics[width=1.00\textwidth,height=0.24\textheight]{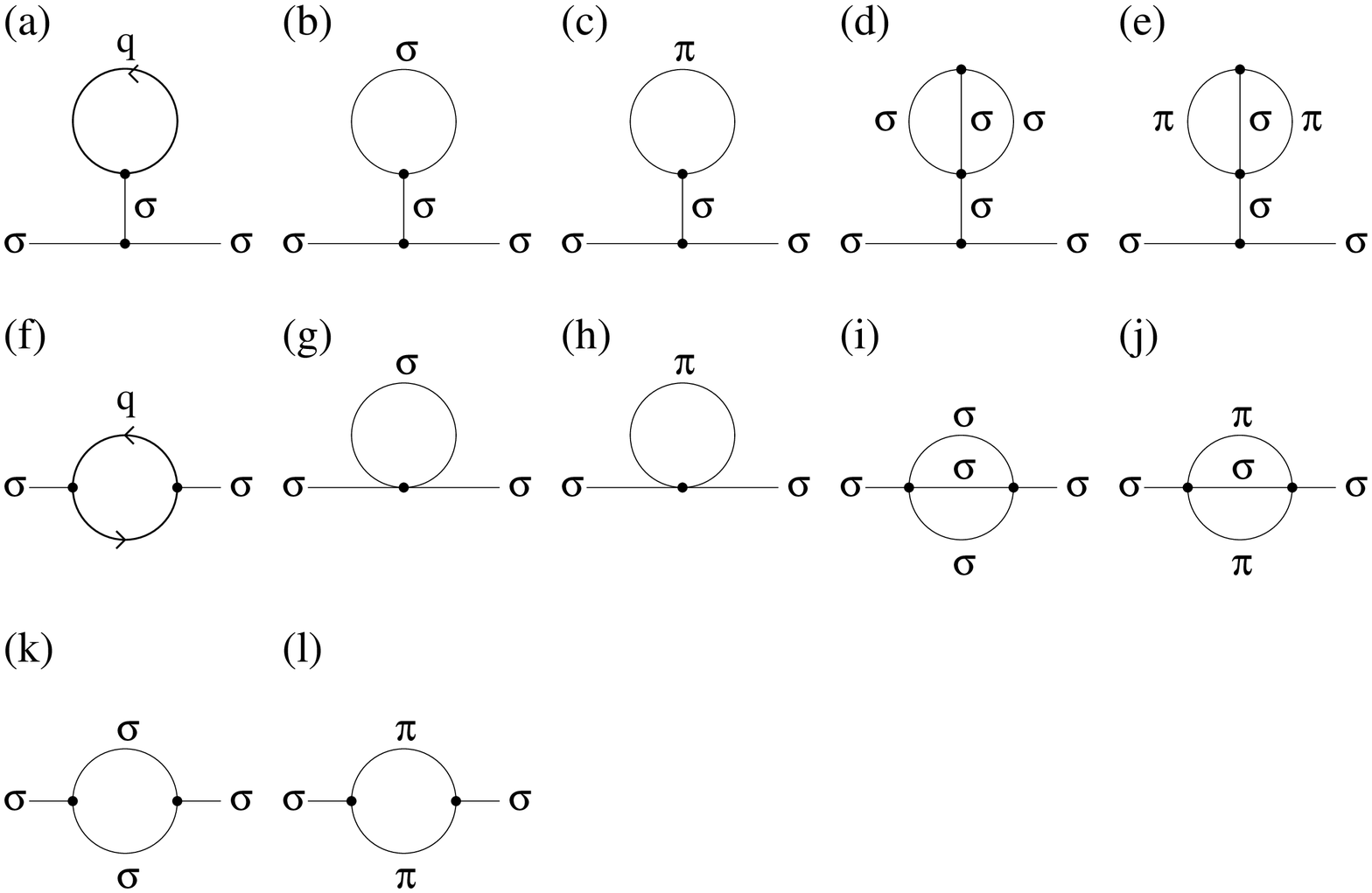}
  \caption{Sigma mass: contributions to the $\sigma$ self-energy} \label{fig3}
\end{figure}

\begin{figure}
  \includegraphics[width=1.00\textwidth,height=0.24\textheight]{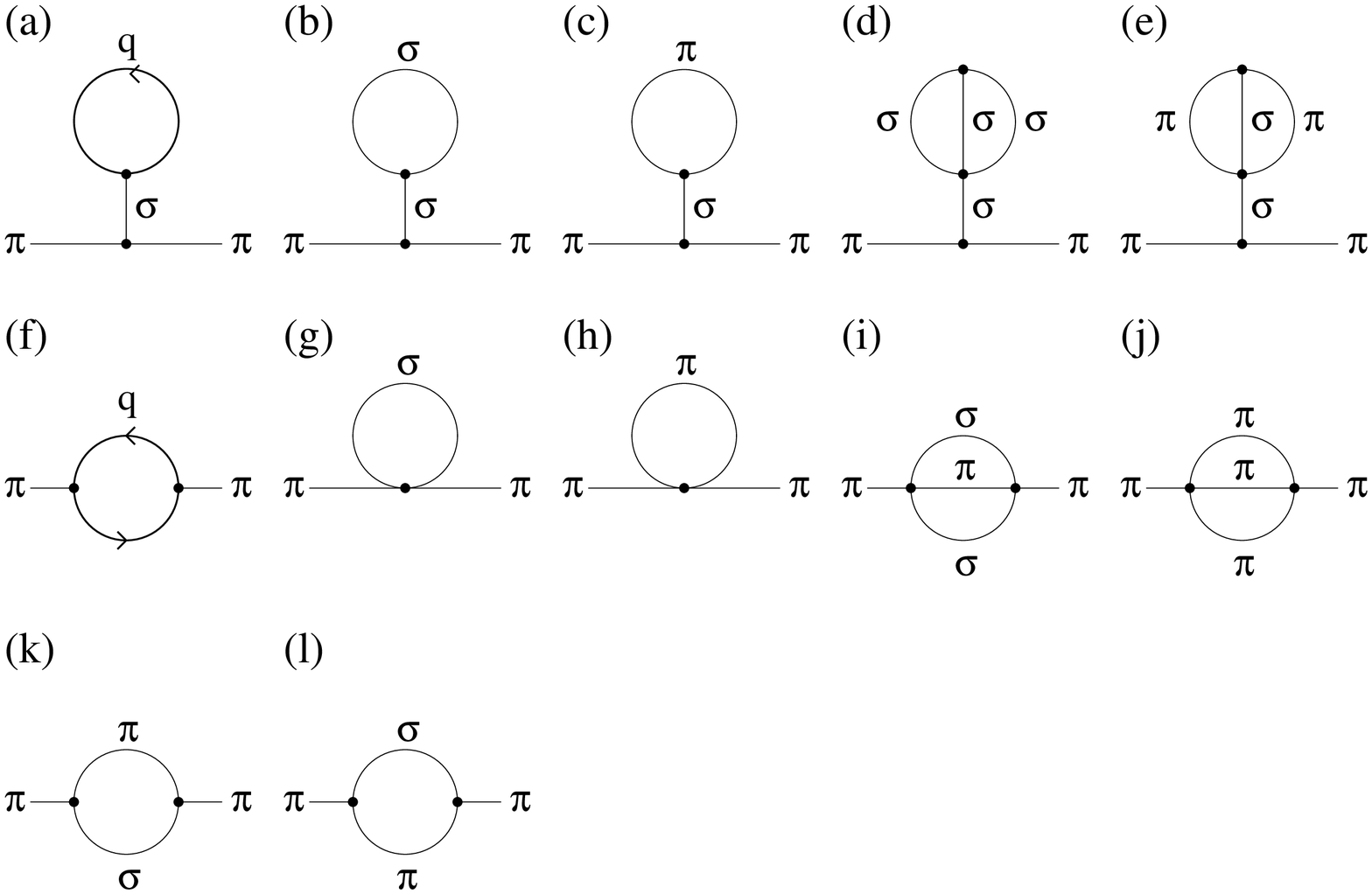}
  \caption{Pion mass: contributions to the $\pi$ self-energy} \label{fig4}
\end{figure}
\clearpage
\noindent \underline{\em \bf Option 2:} {\em dynamical generation following the strategy of Delbourgo and Scadron (DS)}\\[2mm]
In order to illustrate the approach of DS  \cite{Delbourgo:1993dk,Delbourgo:1998kg} we reformulate Eq.\ (\ref{quaddiv4}) slightly:
\begin{eqnarray}
 \lefteqn{S_{(4)}[\vec{\pi}^2] = \frac{i}{2}\int d^4x \;\, \vec{\pi}(x)^2} \nonumber \\[1mm]
 & & \times \Bigg[\int \frac{d^4p}{(2\pi)^4} \; \frac{1}{p^2-\overline{m}^{\,2}} \;  \underbrace{\Bigg\{-\frac{8\,g_{\sigma\pi\pi} \;g\,N_F\,m_q}{m^2_\sigma} + 4\,g^2\,N_F \Bigg\}}_{\mbox{quark contributions $=0$}} \nonumber  \\[1mm]
 & & \;\;\;+ \int \frac{d^4p}{(2\pi)^4} \; \frac{1}{p^2-\overline{m}^{\,2}} \; \underbrace{\Bigg(\frac{2\,g^2_{\sigma\pi\pi}}{m^2_\sigma} -\lambda\Bigg)(3+N_\pi)}_{\mbox{mesonic contributions $=0$}}  \left(1-\frac{\lambda}{4\pi^2}\right)\Bigg] + \ldots  \label{quaddiv6} 
\end{eqnarray}
As indicated above DS set individually the quark and meson contributions to the pion self-energy to zero. The sunset contributions to the pion self-energy being of two-loop order are disregarded implying the replacement $(1-\frac{\lambda}{4\pi^2})\rightarrow 1$ in the expression above. As a result Delbourgo and Scadron obtain the following identities:
\begin{eqnarray} 0 & = & -\frac{8\,g_{\sigma\pi\pi} \;g\,N_F\,m_q}{m^2_\sigma} + 4\,g^2\,N_F \quad \Rightarrow \quad g_{\sigma\pi\pi} = \frac{m^2_\sigma}{2}\frac{g}{m_q} \\[1mm]
0 & = & \frac{2\,g^2_{\sigma\pi\pi}}{m^2_\sigma} -\lambda \quad \Rightarrow \quad \lambda = \frac{2}{m^2_\sigma}\;g^2_{\sigma\pi\pi} = \frac{2}{m^2_\sigma}\;\left(\frac{m^2_\sigma}{2}\frac{g}{m_q}\right)^2 = 2\,g^2 \; \left( \frac{m_\sigma}{2\,m_q}\right)^2 \label{eqnjlds1}
\end{eqnarray}
 The two-loop relation Eq.\ (\ref{eqlambdag1}), i.e.\ $g_{\sigma\pi\pi}= \frac{m_q}{g}\;\lambda$, has been derived already at one-loop order by DS via determination of the following quark-loop contributions to the couplings $g_{\sigma\pi\pi}$ and $\lambda$ performing quark-loop shrinking as demonstrated in Eqs.\ (\ref{eqqloop1}) and (\ref{eqqloop2}):
\begin{eqnarray} g_{\sigma\pi\pi}(\mbox{quark-loop}) & = & -\,4\,g^3\,N_F\,m_q\,i\; I_2(m^2_q) \; \rightarrow \; -\,4\,g^3\,N_F\,m_q\,i \; \frac{i}{16\pi^2} \; =\;  \frac{g^3\,N_F\,m_q}{4\pi^2}\; , \nonumber \\
 & & \\
\lambda(\mbox{quark-loop}) & = & -\,4\,g^4\,N_F\,i\; I_2(m^2_q)\; \quad\;\, \rightarrow \; -\,4\,g^4\,N_F\,i\; \; \frac{i}{16\pi^2} \quad \; =\;  \frac{g^4\,N_F}{4\pi^2} \; .
\end{eqnarray}
The right-hand side of these equations has been obtained here by renormalizing logarithmic divergencies with the help of the log.-divergent gap equation Eq.\ (\ref{eqlogdivgeq1}) choosing the renormalization scale $\overline{m}\approx m_q$. The quark-loop contributions to $m^2_\sigma$ and $m^2_\pi$ obtained by DS are given according to Eqs.\ (\ref{eqeffact3}) and (\ref{eqeffact4}) by
\begin{eqnarray} m^2_\sigma(\mbox{quark-loop})  & = & -\,8\,g^2\,N_F\,m^2_q\,i\; I_2(m^2_q) \; \rightarrow \;-\,8\,g^2\,N_F\,m^2_q\,i \; \frac{i}{16\pi^2} \; =\;  \frac{g^2\,N_F}{8\pi^2} \; (2\,m_q)^2\; , \label{eqmsignjl1} \nonumber \\
 & &  \\
 m^2_\pi(\mbox{quark-loop}) & = & 0 \; . \end{eqnarray}
While focusing on quark-loops the dynamical generation of the SU(2)$\times$SU(2) QLL$\sigma$M performed by DS happens to display results essentially in the Chiral Limit (CL) \cite{vanBeveren:2002mc}. Hence it appears according to DS to be quite appealing and natural that the chiral limiting relation $g_{\sigma\pi\pi}\approx -e^{\,i\,(\alpha-\beta)}\, m^2_\sigma/(2\,|f_\pi|)$ of the L$\sigma$M is resulting directly from assuming the GT relation $m_q\approx -|g| \,f_\pi$ on the quark-level. Moreover is the NJL-relation between $m_\sigma$ and $m_q$ in the CL $|m_\sigma|\approx |2\,m_q|$  \cite{Delbourgo:1982tv} obviously achieved by chosing $|g^2\,N_F/(8\pi^2)|\approx 1$ in Eq.\ (\ref{eqmsignjl1}), or equivalently $|g|=2\pi/\sqrt{N_c}$ with $N_F=2\,N_c=6$, yielding in combination with the GT relation $m_q\approx -|g| \,f_\pi$ and the log.-divergent gap equation Eq.\ (\ref{eqlogdivgeq1}) at a renormalization scale $\overline{m}\approx m_q$ the important correspondence  \cite{Delbourgo:1993dk,Delbourgo:1998kg}
\begin{equation} -\,f_\pi \; \leftrightarrow \; - 2\,N_F \, |g|\; i \int\frac{d^4p}{(2\pi)^4} \; \frac{m_q}{(p^2-m^2_q)^2} \; .
\end{equation}
 
The chiral limiting NJL-relation $|m_\sigma|\approx |2\,m_q|$ implies then in the approach of DS due to Eq.\ (\ref{eqnjlds1}) the relation $\lambda\approx 2 \,|g|^2\;e^{\,2\,i\,(\alpha -\, \beta)}$ between the quartic coupling $\lambda$ and the quark-meson coupling $g$.\\[2mm]
\indent Several comments are here in order:
The very interesting approach of DS has been performed unfortunately just in the limit $e^{\,i\,\alpha}=1$ and $e^{\,i\,(\alpha-\beta)}=-1$ yielding $e^{-\,i\,\beta}=-1$. In this limit the quark-level GT relation reduces to $m_q\approx |g| \,|f_\pi|$ implying the unphysical limit of {\em real-valued} quark masses. Moreover did DS derive the quark-loop contributions to $m_q^2$, $m^2_\pi$ and $m^2_\sigma$ \underline{not} by cancelling systematically quadratic divergencies and renormalizing afterwards the remaining log.-divergent parts of the effective action by making use of the log.-divergent gap equation Eq.\ (\ref{eqlogdivgeq1}). Instead they converted quadratic divergences into logarithmic divergences in making heavy use of Dimensional Regularization (DR) yielding on one hand as a lemma \cite{Delbourgo:1998ji}
\begin{equation} \int \frac{d^4p}{(2\pi)^4} \; \left[ \frac{m^2}{(p^2-m^2)^2} - \, \frac{1}{p^2-m^2} \right] \stackrel{DR}{\longrightarrow} - \frac{i}{16\pi^2} \; m^2 \; ,\end{equation}
on the other hand the vanishing of massless tadpoles, i.e.\ $\int\frac{d^4p}{(2\pi)^4} \frac{1}{p^2} \stackrel{DR}{\longrightarrow} 0$, and the disappearance of quadratic divergences in unrenormalized sunset/sunrise integrals (See e.g.\ also the
DR calculations performed in Refs.\
\cite{Ford:1991hw,Caffo:1998du,Hellman:1986ya}, or on p.\ 114 ff
in Ref.\ \cite{Kleinert:2001ax}). 
From the considerations performed in Option 1 it gets very clear why dimensional regularization \cite{Jack:1989pz,Yang:2001yw,Yang:2002ur} and present implicit regularization \cite{Brizola:1998tn,Sampaio:2002ii} constructed to reproduce results from dimensional regularization are not useful to perform a complete dynamical generation of QLL$\sigma$M-like quantum theories. By erasing massless tadpoles and removing quadratic divergencies in sunset/sunrise diagrams DR does not allow to obtain universal relations between coupling constants like Eqs. (\ref{eqtadpole1}) and (\ref{eqself1}) which permit to cancel quadratic divergencies completely and simultaneously in all parts of the effective action. We would like to add here that similar problems arise also in different regularization schems like Schwinger's regularization \cite{Kleppe:1992qq}, if quadratic divergencies are not treated with due care. In the case of DS the seeming gain in disregarding --- due to the use of DR --- the massless pion-loop tadpole in the tadpole sum contributing to the vacuum expection value of the scalar field spoils e.g.\ the cancellation of quadratic divergencies in the self-energy of the scalar field. Interestingly this problem does not get manifestly visible in the approach of DS as they --- as pointed out above --- convert quadratic into logarithmic divergences with the help of DR. It has to be stressed at this place that the drawbacks of the DR approach of DS {\em do not demerit at all} the great and fascinating implications and insights emerging from the extremely benchmarking work of DS when appreciated correctly.\\[2mm]
\underline{\em\bf Option 3:} {\em dynamical generation of a new standard model of particle physics}\\[2mm]
The most promising way to continue the considerations performed in this manuscript is to apply the strategy outlined in Option 1 to the quantum theory consisting of a sum on one hand of the U(6)$\times$U(6) QLL$\sigma$M (including also vector and axial vector meson fields) to describe the asymptotically free theory of strong interactions and on the other hand the familiar theory of electroweak interactions. In such an approach the universal relations between coupling constants like Eqs. (\ref{eqtadpole1}) and (\ref{eqself1}) to be used to cancel quadratic divergences would extend of course to all parts of the effective action stemming from the strong {\em and} electroweak sector of the whole theory. The quarks and antiquarks in such an approach should be of course considered as constituent quarks rather than current quarks. A small source of mixing between scalar and pseudoscalar fields in such a theory should take care of experimental facts related e.g.\ to chiral symmetry breaking \cite{Scadron:2006mq} and flavour changing neutral currents like scalar meson dominance \cite{Kleefeld:2005at,Schechter:1993tc} in semileptronic and non-leptonic meson decays.   

As we have noted already above the cancellation of quadratic divergences in such a theory is interconnecting different loop orders of the theory. This feature typically known from non-Abelian gauge theories is recovered in this manuscript already for a theory containing merely spin 1/2 Fermions and scalar and pseudoscalar spin 0 Bosons. Moreover would we like to point out that a complete knowledge of the one-point function of the Higgs field to be obtained in the same way as described here for the one-point function of the $\sigma$-meson in the SU(2)$\times$SU(2) QLL$\sigma$M without vector and axial vector meson fields would allow to make a theoretical prediction for the vacuum expectation value of the Higgs   on the basis of the knowlege of the finite part of the sunset/sunrise integral with zero external four-momentum \cite{Yang:2003bv}.\footnote{Working out this manuscript we noted a sign mistake in the  quadratically divergent term of the sunset integral in the article of Inami {\em et al.} \cite{Deshpande:1983ka} which made us to invoke the wrong sign in our considations on the sunset integral in footnote 6 of Ref.\ \cite{Kleefeld:2005hd}. Applying the corrected expressions of Appendix B to our discussion of the 1-point function of the $\phi^4$-theory in Ref.\ \cite{Kleefeld:2005hd} we obtain a corrected Eq.\ (18) of Ref.\ \cite{Kleefeld:2005hd}:\\
\mbox{} $\;\;\;S_{{}_{(1)}}[\,\phi\,] = \int d^4z \, \left( - \frac{1}{3!} \;g_{{}_{(1)}} \right)  3\,i \; \phi_{{}_{(1)}}(z) \; \Big\{ \left( 1-\frac{2}{3}\, \frac{1}{16\pi^2} \, \lambda_{{}_{(1)}}\right)  I_1(m^2_{{}_{(1)}})  -i\, \left( \frac{1}{16\pi^2} \right)^2  m^2_{{}_{(1)}}  \frac{(8 -  C)}{3} \,\lambda_{{}_{(1)}} \Big\} +  \ldots\,. \;\;\;(18)$\\
Hence the non-trivial cancellation of quadratic divergencies yields $\lambda_{{}_{(1)}}=+(3/2)\,
16\pi^2=+24\pi^2$ implying due to $\lambda_{{}_{(1)}} =  - g^4_{{}_{(1)}}/(32\pi^2 m^4_{{}_{(1)}})$ now $g_{{}_{(1)}}=\pm 4\pi \;(+i)^{1/2}\, 3^{1/4} \, m_{{}_{(1)}}$ and
$g_{{}_{(1)}}=\pm  4\pi \;(-i)^{1/2}\, 3^{1/4} \, m_{{}_{(1)}}$.}

{\bf Acknowledgments.} It has been a great pleasure and honour to collaborate with M.D.~Scadron over many years. With this manuscript we would like to deliver our very best wishes to Mike and Arlene on the occasion of Mike's 70th birthday on February 12, 2008. This work has been supported by the
FCT of the {\em Minist\'{e}rio da Ci\^{e}ncia, Tecnologia e Ensino Superior} \/of Portugal, under Grants no.\ PDCT/FP/63907/2005, POCI/FP/81913/2007 and the Doppler and Nuclear Physics Institute (Dep.~Theor.~Phys.) at the Academy of Sciences of the Czech Republic by Project no.\ LC06002.

\clearpage
\begin{appendix}
\section{A. List of important integrals}
We want to list here some important integral identities to be used in the manuscript:\footnote{In all integrals we assume the imaginary part of the squared masses to be negative.}
\begin{eqnarray} I_n(m^2) & \equiv & \int \frac{d^4p}{(2\pi)^4} \; \frac{1}{(p^2-m^2)^n} \; \stackrel{n\ge 3}{=}\; (-1)^n\, \frac{i}{16\,\pi^2}\; \frac{1}{(n-1)!\; m^{2n-4}} \;  ,\label{eqint1} \quad \\
 I_1(m^2) & \equiv & \int \frac{d^4p}{(2\,\pi)^4}\, \frac{1}{p^2-m^2} \; ,  \\
 I_2(m^2) & \equiv & \int \frac{d^4p}{(2\,\pi)^4}\, \frac{1}{(p^2-m^2)^2} \; ,  \\
 I_3(m^2) & = & \int \frac{d^4p}{(2\,\pi)^4}\, \frac{1}{(p^2-m^2)^3} \;  = \; -\, \frac{i}{32\,\pi^2}\; \frac{1}{m^2} \; ,  \\
 I_4(m^2) & = & \int \frac{d^4p}{(2\,\pi)^4}\, \frac{1}{(p^2-m^2)^4}\; = \; \frac{1}{3}\, \frac{dI_3(m^2)}{dm^2} \;  = \; + \frac{i}{96\,\pi^2}\; \frac{1}{m^4} \; ,  \\
 I_{1,1}(m^2_1,m^2_2) & \equiv & \int \frac{d^4p}{(2\,\pi)^4}\, \frac{1}{(p^2-m^2_1)(p^2-m^2_2)} \; ,  \\[2mm]
 I_{2,1}(m^2_1,m^2_2) & \equiv & \int \frac{d^4p}{(2\,\pi)^4}\, \frac{1}{(p^2-m^2_1)^2(p^2-m^2_2)} \nonumber \\[2mm]
 & = & \frac{i}{16\,\pi^2}\; \frac{1}{(m^2_2-m^2_1)} \; \left( 1- \frac{m^2_2}{m^2_2-m^2_1} \; \ln \frac{m_2^2}{m^2_1} \right) \; , \\[1mm]
I_{1,1,1}(m^2_1,m^2_2,m^2_3) & \equiv & \int \frac{d^4p}{(2\,\pi)^4}\, \frac{1}{(p^2-m^2_1)(p^2-m^2_2)(p^2-m^2_3)} \nonumber \\
 & = & \frac{i}{16\,\pi^2}\; \frac{\displaystyle m^2_1\,m^2_2 \; \ln \frac{m_1^2}{m^2_2} + m^2_2\,m^2_3 \; \ln \frac{m_2^2}{m^2_3} + m^2_3\,m^2_1 \; \ln \frac{m_3^2}{m^2_1}}{\displaystyle (m_1^2-m^2_2)(m_2^2-m^2_3)(m_3^2-m^2_1)} \; ,  \\[1mm]
 I_{2,2}(m^2_1,m^2_2) & \equiv &\int \frac{d^4p}{(2\,\pi)^4}\, \frac{1}{(p^2-m^2_1)^2(p^2-m^2_2)^2} \nonumber \\
 & = & \frac{d}{dm^2_2}\int \frac{d^4p}{(2\,\pi)^4}\, \frac{1}{(p^2-m^2_1)^2(p^2-m^2_2)} \nonumber \\
 & = & \frac{d}{dm^2_2} \left( \frac{i}{16\,\pi^2}\; \frac{1}{(m^2_2-m^2_1)} \; \left( 1- \frac{m^2_2}{m^2_2-m^2_1} \; \ln \frac{m_2^2}{m^2_1} \right) \right) \nonumber \\[2mm]
 & = & \frac{i}{16\,\pi^2}\; \frac{1}{(m^2_2-m^2_1)^2} \; \left( -\,2+ \frac{m^2_1+m^2_2}{m^2_2-m^2_1} \; \ln \frac{m_2^2}{m^2_1} \right) \; .
\end{eqnarray}
\clearpage
\section{B. The sunset / sunrise diagram}
The leading divergent part of the sunset diagram for zero external four-momentum and equal masses has been determined in cutoff regularization by Ji-Feng Yang, Jie Zhou and Chen Wu to be \cite{Yang:2003bv}:\footnote{For a discussion of the finite part of the sunset/sunrise integral for non-zero external four-momentum
on the basis of implicit renormalization see e.g.\ Ref.\
\cite{Sampaio:2002ii}.}
\begin{eqnarray} \lefteqn{I^{\,\Lambda}_{sunset}(m^2)= \int^\Lambda \frac{d^4p_1}{(2\pi)^4} \;\int^\Lambda \frac{d^4p_2}{(2\pi)^4} \;\int^\Lambda \frac{d^4p_3}{(2\pi)^4} \; \; \frac{(2\pi)^4\; \delta^4 (p_1+p_2+p_3)}{(p_1^2- m^2)(p_2^2- m^2)(p_3^2- m^2)}\; =} \nonumber \\[1mm]
 & = & \left( \frac{1}{16\,\pi^2} \right)^2 \, \left( 2\; \Lambda^2 - \frac{3}{2}\; m^2\; \ln^2\left( \frac{\Lambda^2}{m^2} \right) - \, 3\, m^2\; \ln\left( \frac{\Lambda^2}{m^2} \right) + C\, m^2 \right)  + O(\Lambda^{-2})\; ,\quad
\end{eqnarray}  
while the integration constant $C$ was numerically estimated by Ji-Feng Yang {\em et al.}\ to be approximately $C\simeq 4$. A numerical analysis by G.\ Rupp (private communicaton, 22.05.2006) yields $C\in [4.160805,4.160810]$. The above result can be slightly rewritten:
\begin{eqnarray} \lefteqn{I^{\,\Lambda}_{sunset}(m^2)= \int^\Lambda \frac{d^4p_1}{(2\pi)^4} \;\int^\Lambda \frac{d^4p_2}{(2\pi)^4} \;\int^\Lambda \frac{d^4p_3}{(2\pi)^4} \; \; \frac{(2\pi)^4\; \delta^4 (p_1+p_2+p_3)}{(p_1^2- m^2)(p_2^2- m^2)(p_3^2- m^2)}\; =} \nonumber \\[1mm]
 & = & \left( \frac{1}{16\,\pi^2} \right)^2 \, \Bigg(  2\, \left( \Lambda^2 - m^2 \; \ln \left( \frac{\Lambda^2}{m^2}\right) \right) - \frac{3}{2}\; m^2  \left( \ln\left( \frac{\Lambda^2}{m^2} \right) - \, 1\right)^2 \nonumber \\
 &  & \qquad \qquad\quad -  4\, m^2 \, \left( \ln\left( \frac{\Lambda^2}{m^2}\right) \, - \, 1 \right) + m^2 \, \left(C-\frac{5}{2}\right) \Bigg)  + O(\Lambda^{-2}) \nonumber \\
 & \rightarrow &  2\, \frac{i}{16\,\pi^2} \,I_1(m^2) + \frac{3}{2}\; m^2 \; I_2(m^2)^2 + \, 4 \; m^2  \frac{i}{16\,\pi^2} \; I_2(m^2) -\, \left( \frac{i}{16\,\pi^2} \right)^2 \; m^2 \, \left(C-\frac{5}{2}\right) \; .\nonumber \\ 
\end{eqnarray}  
The last line displays manifestly the most divergent part of the massive sunset diagram at zero external four-momentum in an regularization sheme independent manner.

Now we may apply the renormalization procedure of Delbourgo and Scadron replacing the log.-divergent Bosonic one-loop integral at some renormalization scale $\overline{m}$ ($\simeq \hat{m}= m_q$) by the finite number $+\; \frac{i}{16\, \pi^2}$ (by adding a suitable counterterm), i.e. $I_2(\overline{m}^{\,2})\equiv \int \frac{d^4p}{(2\pi)^4} \;  \frac{1}{(p^2-\overline{m}^{\,2})^2}  \; \rightarrow \; +\; \frac{i}{16\, \pi^2}$ 
being known as {\em log.-divergent gap equation}. Then the sunset integral in regularization sheme indepent representation reduces to:
\begin{eqnarray} I_{sunset}(\overline{m}^{\,2}) & =& \int \frac{d^4p_1}{(2\pi)^4} \;\int \frac{d^4p_2}{(2\pi)^4} \;\int \frac{d^4p_3}{(2\pi)^4} \; \; \frac{(2\pi)^4\; \delta^4 (p_1+p_2+p_3)}{(p_1^2-\overline{m}^{\,2})(p_2^2-\overline{m}^{\,2})(p_3^2-\overline{m}^{\,2})} \nonumber \\[2mm]
 & \rightarrow &   2\; \frac{i}{16\,\pi^2} \; I_1(\overline{m}^{\,2})\;  -\, \left( \frac{1}{16\,\pi^2} \right)^2 \; \overline{m}^{\,2} \, (8- \; C)  \; .
\end{eqnarray}
The situation is more involved if the equal mass sunset integral isn't evaluated at the renormalization scale $\overline{m}$, yet at some arbitrary mass $m$. Using Eqs.\ (\ref{eqid1}) and (\ref{eqid2}) we obtain:
\begin{eqnarray} \lefteqn{I_{sunset}(m^2) = \int \frac{d^4p_1}{(2\pi)^4} \;\int \frac{d^4p_2}{(2\pi)^4} \;\int \frac{d^4p_3}{(2\pi)^4} \; \; \frac{(2\pi)^4\; \delta^4 (p_1+p_2+p_3)}{(p_1^2-m^2)(p_2^2-m^2)(p_3^2-m^2)}} \nonumber \\[2mm]
 & \rightarrow &  2\, \frac{i}{16\,\pi^2} \,I_1(m^2) + \frac{3}{2}\; m^2 \; I_2(m^2)^2 + \, 4 \; m^2  \frac{i}{16\,\pi^2} \; I_2(m^2) -\, \left( \frac{i}{16\,\pi^2} \right)^2 \; m^2 \, \left(C-\frac{5}{2}\right) \nonumber \\[2mm]
 & \rightarrow &  2\, \frac{i}{16\,\pi^2} \,\left( I_1(\overline{m}^{\,2}) +\frac{i}{16\,\pi^2}\;(m^2-\overline{m}^{\,2}) \; \left( 2- \frac{m^2}{m^2-\overline{m}^{\,2}} \; \ln \frac{m^2}{\overline{m}^{\,2}} \right) \right)  \nonumber \\[2mm]
 & & + \frac{3}{2}\; m^2 \; \left(\frac{i}{16\,\pi^2}\right)^2 \left( 1 - \,\ln \frac{m^2}{\overline{m}^{\,2}}\right)^2 + \, 4 \; m^2 \left( \frac{i}{16\,\pi^2}\right)^2  \left( 1 - \,\ln \frac{m^2}{\overline{m}^{\,2}}\right) \nonumber \\
 & & + \left( \frac{i}{16\,\pi^2} \right)^2 \; m^2 \, \left(\frac{5}{2}-C\right) \nonumber \\[2mm]
 & = &   \frac{i}{16\,\pi^2} \,\Bigg\{2\, I_1(\overline{m}^{\,2})  +   \frac{i}{16\,\pi^2} \; \left(  \frac{3}{2}\; m^2\,  \left(\ln \frac{m^2}{\overline{m}^{\,2}}\right)^2 -9\,m^2\,\ln \frac{m^2}{\overline{m}^{\,2}}  +  m^2\,(12 -C) -4\,\overline{m}^{\,2} \right)\Bigg\} .\nonumber \\ 
\end{eqnarray}
In order to complete our discussion it would be necessary to consider some special cases of the more general class of sunset-like integrals with zero external four-momentum of the following type:
\begin{eqnarray} \lefteqn{I^{\,sunset}_{n_1,n_2,n_3}(m_1^2,m_2^2,m^2_3) \equiv} \nonumber \\
 & \equiv & \int \frac{d^4p_1}{(2\pi)^4} \;\int \frac{d^4p_2}{(2\pi)^4} \;\int \frac{d^4p_3}{(2\pi)^4} \; \; \frac{(2\pi)^4\; \delta^4 (p_1+p_2+p_3)}{(p_1^2-m_1^2)^{n_1}(p_2^2-m_2^2)^{n_2}(p_3^2-m_3^2)^{n_3}}
\end{eqnarray}
yielding of course as a special case also the above discussed integral $I_{sunset}(m^2)=I^{\,sunset}_{1,1,1}(m^2,m^2,m^2)$. 
Unfortunately there is lacking yet an analysis of the  the quadratically divergent integral $I^{\,sunset}_{1,1,1}(m^{\prime \, 2},m^2,m^2)$ beyond its quadratic divergence in the same spirit as it has been provided above for the integral $I_{\,sunset}(m^2)$:
\begin{eqnarray} \lefteqn{I^{\,sunset}_{1,1,1}(m^{\prime \, 2},m^2,m^2)=} \nonumber \\[2mm]
 & = & \int \frac{d^4p_1}{(2\pi)^4} \;\int \frac{d^4p_2}{(2\pi)^4} \;\int \frac{d^4p_3}{(2\pi)^4} \; \; \frac{(2\pi)^4\; \delta^4 (p_1+p_2+p_3)}{(p_1^2-m^{\prime\,2})(p_2^2-m^2)(p_3^2-m^2)} \nonumber \\[2mm]
 & = & I_{\,sunset}(m^2) + (m^{\prime\,2}-m^2) \times \nonumber \\[2mm]
 & & \times \int \frac{d^4p_1}{(2\pi)^4} \;\int \frac{d^4p_2}{(2\pi)^4} \;\int \frac{d^4p_3}{(2\pi)^4} \; \; \frac{(2\pi)^4\; \delta^4 (p_1+p_2+p_3)}{(p_1^2-m^{\prime\,2})(p_1^2-m^2)(p_2^2-m^2)(p_3^2-m^2)} \; .\quad
\end{eqnarray}
As it can be seen by inspection of Appendix C the integral $I^{\,sunset}_{1,1,1}(m^{\prime \, 2},m^2,m^2)$ appears at various places in the effective actions under consideration.
\clearpage
\section{C. Effective actions} 
The effective action $S_{(1)}[\sigma]$ of the scalar one-point function consisting of the contributions illustrated in Fig.\ \ref{fig1} is given by \cite{Kleefeld:2005hd}:
\begin{eqnarray}  \lefteqn{S_{(1)}[\sigma] =S_{(1a)}[\sigma] +S_{(1b)}[\sigma] +S_{(1c)}[\sigma] +S_{(1d)}[\sigma] +S_{(1e)}[\sigma]} \nonumber \\[1mm]
 & = & \int d^4x \; \sigma(x) \Big\{  \left<0\right|T\Big[ g \; \overline{q^c_+}(x)q_-(x)+ 3\,g_{\sigma\pi\pi}\,\sigma(x)^2+N_\pi \,g_{\sigma\pi\pi}\,\pi(x)^2 \Big]\left|0\right> \nonumber \\
 & & \qquad \qquad\;\; + \,i \int d^4x^\prime \, \left( - \frac{\lambda}{4}\right)\,g_{\sigma\pi\pi}\left<0\right|T[\sigma(x)\,\sigma(x^\prime)]\left|0\right>\nonumber \\
 & & \qquad \qquad\;\; \times\,8\,\Big(3\,\left<0\right|T[\sigma(x)\,\sigma(x^\prime)]\left|0\right>^2+N_\pi\,\left<0\right|T[\pi(x)\,\pi(x^\prime)] \left|0\right>^2 \Big)\Big\}\nonumber \\
 & = & \int d^4x \; \sigma(x) \Big\{ i\int \frac{d^4p}{(2\pi)^4} \left( - \, \frac{4\,g \,N_F\,m_q}{p^2-m_q^2}+ \frac{3\,g_{\sigma\pi\pi}}{p^2-m^2_\sigma}+\frac{N_\pi \,g_{\sigma\pi\pi}}{p^2-m^2_\pi} \right) \nonumber \\
 & & \qquad \qquad\;\; + \,i \int d^4x^\prime\; i^3\int \frac{d^4p_1}{(2\pi)^4}\int \frac{d^4p_2}{(2\pi)^4}\int \frac{d^4p_3}{(2\pi)^4}\; e^{-i(p_1+p_2+p_3)\cdot(x-x^\prime)}\nonumber \\
 & & \qquad \qquad\;\; \times\,\left( - \frac{\lambda}{4}\right)\, \frac{8\,g_{\sigma\pi\pi}}{p^2_1-m^2_\sigma}\,\left(\frac{3}{(p^2_2-m^2_\sigma)(p^2_3-m^2_\sigma)}+\frac{N_\pi}{(p^2_2-m^2_\pi)(p^2_3-m^2_\pi)}\right)\Big\}\nonumber \\[1mm]
 & = & \int d^4x \; \sigma(x)\; i\,\Big\{ \int \frac{d^4p}{(2\pi)^4} \left( - \, \frac{4\,g \,N_F\,m_q}{p^2-m_q^2}+ \frac{3\,g_{\sigma\pi\pi}}{p^2-m^2_\sigma}+\frac{N_\pi \,g_{\sigma\pi\pi}}{p^2-m^2_\pi} \right) \nonumber \\
 & & \qquad \qquad\;\; +\,i  \int \frac{d^4p_1}{(2\pi)^4}\int \frac{d^4p_2}{(2\pi)^4}\int \frac{d^4p_3}{(2\pi)^4}\; (2\pi)^4\, \delta^4(p_1+p_2+p_3)\nonumber \\
 & & \qquad \qquad\;\; \times\,\, \frac{2\,\lambda\,g_{\sigma\pi\pi}}{p^2_1-m^2_\sigma}\,\left(\frac{3}{(p^2_2-m^2_\sigma)(p^2_3-m^2_\sigma)}+\frac{N_\pi}{(p^2_2-m^2_\pi)(p^2_3-m^2_\pi)}\right)\Big\} \nonumber \\[2mm]
 & + & \mbox{non-local terms} \nonumber \\[3mm]
 & = & \int d^4x \; \sigma(x)\; i\,\Big\{  - \, 4\,g \,N_F\,m_q \; I_1(m^2_q) + 3\,g_{\sigma\pi\pi}\;I_1(m^2_\sigma)+N_\pi \,g_{\sigma\pi\pi}\;I_1(m^2_\pi) \nonumber \\
 & & \qquad \qquad\quad \;\;\;\;\, +\, 2\,\lambda\,g_{\sigma\pi\pi} \; i\,\Big(3\;I^{\,sunset}_{1,1,1}(m^2_\sigma,m^2_\sigma,m^2_\sigma)+N_\pi\;I^{\,sunset}_{1,1,1}(m^2_\sigma,m^2_\pi,m^2_\pi)\Big)\Big\} \nonumber \\[2mm]
 & + & \mbox{non-local terms} \; .
\end{eqnarray}
The effective action $S_{(2)}[\bar{q}q]$ for the two-point function of the (anti)quarks consisting of the contributions illustrated in Fig.\ \ref{fig2} is given by  \cite{Kleefeld:2005hd}:
\begin{eqnarray} \lefteqn{S_{(2)}[\bar{q}q]=} \nonumber \\[2mm]
 & = &  S_{(2a)}[\bar{q}q] +S_{(2b)}[\bar{q}q] +S_{(2c)}[\bar{q}q] +S_{(2d)}[\bar{q}q] +S_{(2e)}[\bar{q}q]+S_{(2f)}[\bar{q}q] +S_{(2g)}[\bar{q}q] \nonumber \\[4mm]
 & = & \frac{i}{2}\int d^4x \; \overline{q^c_+}(x)\,q_-(x) \;\int d^4z \;   \left<0\right|T[\sigma(x)\,\sigma(z)]\left|0\right>\nonumber \\
 & & \quad \quad\, \times\, 2\,g \;\Big\{  \left<0\right|T\Big[ g \; \overline{q^c_+}(z)\,q_-(z)+ 3\,g_{\sigma\pi\pi}\,\sigma(z)^2+N_\pi \,g_{\sigma\pi\pi}\,\pi(z)^2 \Big]\left|0\right> \nonumber \\
 & & \qquad \qquad\;\; + \,i \int d^4z^\prime \, \left( - \frac{\lambda}{4}\right)\,g_{\sigma\pi\pi}\left<0\right|T[\sigma(z)\,\sigma(z^\prime)]\left|0\right>\nonumber \\
 & & \qquad \qquad\;\; \times\,8\,\Big(3\,\left<0\right|T[\sigma(z)\,\sigma(z^\prime)]\left|0\right>^2+N_\pi\,\left<0\right|T[\pi(z)\,\pi(z^\prime)] \left|0\right>^2 \Big)\Big\} \nonumber \\
 & + & \frac{i}{2}\int d^4x \int d^4x^\prime \;2\,g^2 \;\Big\{  \overline{q^c_+}(x)\,\left<0\right|T\Big[ q_-(x)\,\overline{q^c_+}(x^\prime) \Big]\left|0\right> \,q_-(x^\prime)\; \left<0\right|T[\sigma(x)\,\sigma(x^\prime)]\left|0\right>\nonumber \\
 & &  \qquad\qquad + N_\pi \; \overline{q^c_+}(x)\,i\,\gamma_5 \, \left<0\right|T\Big[  q_-(x)\,\overline{q^c_+}(x^\prime)\Big]\left|0\right> \,i\,\gamma_5 \,\,q_-(x^\prime) \; \left<0\right|T[\pi(x)\,\pi(x^\prime)]\left|0\right>\Big\} \nonumber \\
 & = & \frac{i}{2}\int d^4x \;\, \overline{q^c_+}(x)\,q_-(x) \;\int d^4z \;\,i \int \frac{d^4q}{(2\pi)^4} \; e^{-i q\cdot (x-z)}  \frac{1}{q^2-m^2_\sigma}\nonumber \\
 & & \quad \quad\, \times\,2\, g \;\Big\{  i \int \frac{d^4p}{(2\pi)^4} \;\left( -\frac{4\,g\,N_F\,m_q}{p^2-m^2_q} +\frac{3\,g_{\sigma\pi\pi}}{p^2-m_\sigma^2}+\frac{N_\pi \,g_{\sigma\pi\pi}}{p^2-m^2_\pi} \right) \nonumber \\
 & & \qquad \qquad\;\; + \,i \int d^4z^\prime \; i^3\int \frac{d^4p_1}{(2\pi)^4}\int \frac{d^4p_2}{(2\pi)^4}\int \frac{d^4p_3}{(2\pi)^4}\; e^{-i(p_1+p_2+p_3)\cdot(z-z^\prime)}\nonumber \\
 & & \qquad \qquad\;\; \times\,\left( - \frac{\lambda}{4}\right) \frac{8\,g_{\sigma\pi\pi}}{p^2_1-m^2_\sigma}\,\,\left(\frac{3}{(p^2_2-m^2_\sigma)(p^2_3-m^2_\sigma)}+\frac{N_\pi}{(p^2_2-m^2_\pi)(p^2_3-m^2_\pi)}\right)\Big\} \nonumber \\
 & + & \frac{i}{2}\int d^4x \int d^4x^\prime\; i^2\int \frac{d^4p_1}{(2\pi)^4}\int \frac{d^4p_2}{(2\pi)^4} \;\; e^{-i(p_1+p_2)\cdot(x-x^\prime)}\nonumber \\
 & &  \qquad\qquad \times 2\,g^2 \;\overline{q^c_+}(x)\;\left(  \frac{(\not\!\!p_1 + m_q)}{(p^2_1-m^2_q)(p^2_2-m^2_\sigma)}+ \frac{N_\pi\,(\not\!\!p_1 - m_q)}{(p^2_1-m^2_q)(p^2_2-m^2_\pi)}\right)\,q_-(x^\prime) \nonumber \\
 & = & \frac{i}{2}\int d^4x \;\, \overline{q^c_+}(x)\,q_-(x)\; \frac{2\,g}{m^2_\sigma} \;\Big\{ \int \frac{d^4p}{(2\pi)^4} \;\left( -\frac{4\,g\,N_F\,m_q}{p^2-m^2_q} +\frac{3\,g_{\sigma\pi\pi}}{p^2-m_\sigma^2}+\frac{N_\pi \,g_{\sigma\pi\pi}}{p^2-m^2_\pi} \right) \nonumber \\
 & & \qquad \qquad\qquad \qquad\quad\;\; + \,i \int \frac{d^4p_1}{(2\pi)^4}\int \frac{d^4p_2}{(2\pi)^4}\int \frac{d^4p_3}{(2\pi)^4}\; (2\pi)^4\, \delta^4(p_1+p_2+p_3)\nonumber \\
 & & \qquad \qquad\qquad \qquad\quad\;\; \times\, \frac{2\,\lambda\,g_{\sigma\pi\pi}}{p^2_1-m^2_\sigma}\,\,\left(\frac{3}{(p^2_2-m^2_\sigma)(p^2_3-m^2_\sigma)}+\frac{N_\pi}{(p^2_2-m^2_\pi)(p^2_3-m^2_\pi)}\right)\Big\} \nonumber \\[1mm]
 & - & \frac{i}{2}\int d^4x \;\,\overline{q^c_+}(x)\;q_-(x)  \; 2\,g^2\;m_q \int \frac{d^4p}{(2\pi)^4}\left(  \frac{1}{(p^2-m^2_q)(p^2-m^2_\sigma)}- \frac{N_\pi}{(p^2-m^2_q)(p^2-m^2_\pi)}\right) \nonumber \\[1mm]
 & + & \mbox{non-local terms}\nonumber \\
 & = & \frac{i}{2}\int d^4x \;\, \overline{q^c_+}(x)\,q_-(x)\; \frac{2\,g}{m^2_\sigma} \;\Big\{  - \, 4\,g \,N_F\,m_q \; I_1(m^2_q) + 3\,g_{\sigma\pi\pi}\;I_1(m^2_\sigma)+N_\pi \,g_{\sigma\pi\pi}\;I_1(m^2_\pi) \nonumber \\
 & & \qquad \qquad\qquad\qquad \;\;\;\;\, +\, 2\,\lambda\,g_{\sigma\pi\pi} \; i\,\Big(3\;I^{\,sunset}_{1,1,1}(m^2_\sigma,m^2_\sigma,m^2_\sigma)+N_\pi\;I^{\,sunset}_{1,1,1}(m^2_\sigma,m^2_\pi,m^2_\pi)\Big)\Big\} \nonumber \\[1mm]
 & - & \frac{i}{2}\int d^4x \;\,\overline{q^c_+}(x)\;q_-(x)  \; 2\,g^2\;m_q \;\Big(  I_{1,1}(m^2_q,m^2_\sigma)- \,N_\pi\; I_{1,1}(m^2_q,m^2_\pi) \Big) \nonumber \\[1mm]
 & + & \mbox{non-local terms} \; .
\end{eqnarray}
The effective action $S_{(3)}[\sigma^2]$ for the two-point function of the $\sigma$ consisting of the contributions illustrated in Fig.\ \ref{fig3} is given by  \cite{Kleefeld:2005hd}:
\begin{eqnarray} \lefteqn{S_{(3)}[\sigma^2]=} \nonumber \\[2mm]
 & + & S_{(3a)}[\sigma^2] +S_{(3b)}[\sigma^2] +S_{(3c)}[\sigma^2] +S_{(3d)}[\sigma^2] +S_{(3e)}[\sigma^2] \nonumber \\
 & + & S_{(3f)}[\sigma^2] +S_{(3g)}[\sigma^2] +S_{(3h)}[\sigma^2] +S_{(3i)}[\sigma^2] +S_{(3j)}[\sigma^2]+S_{(3k)}[\sigma^2] +S_{(3l)}[\sigma^2] \nonumber \\[1mm]
 & = & \frac{i}{2}\int d^4x \; \sigma(x)^2 \;\int d^4z \;   \left<0\right|T[\sigma(x)\,\sigma(z)]\left|0\right>\nonumber \\
 & & \quad \quad\,\;\; \times\, 6\,g_{\sigma\pi\pi}\, \Big\{  \left<0\right|T\Big[ g \; \overline{q^c_+}(z)\,q_-(z)+ 3\,g_{\sigma\pi\pi}\,\sigma(z)^2+N_\pi \,g_{\sigma\pi\pi}\,\pi(z)^2 \Big]\left|0\right> \nonumber \\
 & & \qquad \qquad\qquad\;\; + \,i \int d^4z^\prime \, \left( - \frac{\lambda}{4}\right)\,g_{\sigma\pi\pi}\left<0\right|T[\sigma(z)\,\sigma(z^\prime)]\left|0\right>\nonumber \\
 & & \qquad \qquad\qquad\;\; \times\,8\,\Big(3\,\left<0\right|T[\sigma(z)\,\sigma(z^\prime)]\left|0\right>^2+N_\pi\,\left<0\right|T[\pi(z)\,\pi(z^\prime)] \left|0\right>^2 \Big)\Big\} \nonumber \\[1mm]
 & + & \frac{i}{2}\int d^4x \int d^4x^\prime \;\sigma(x)\,\sigma(x^\prime)\,g^2 \left<0\right|T\Big[ \overline{q^c_+}(x)\,q_-(x)\;\overline{q^c_+}(x^\prime)\,q_-(x^\prime) \Big]\left|0\right>_c  \nonumber \\[1mm]
 & + & \int d^4x \;\sigma(x)^2 \;\, 2\,\left(-\,\frac{\lambda}{4}\right)\, \Big( 3\,\left<0\right|T[\sigma(x)\,\sigma(x)]\left|0\right> + N_\pi \,\left<0\right|T[\pi(x)\,\pi(x)]\left|0\right>\Big)\nonumber \\[1mm]
 & + & \frac{i}{2}\int d^4x \; \int d^4x^\prime \; \sigma(x)\,\sigma(x^\prime) \;  \left<0\right|T[\sigma(x)\,\sigma(x^\prime)]\left|0\right>\nonumber \\
 & & \qquad \qquad\qquad\;\; \times\,2\,\lambda^2\,\Big(3\,\left<0\right|T[\sigma(x)\,\sigma(x^\prime)]\left|0\right>^2+N_\pi\,\left<0\right|T[\pi(x)\,\pi(x^\prime)] \left|0\right>^2 \Big) \nonumber \\[1mm]
 & + & \frac{i}{2}\int d^4x \int d^4x^\prime \; \sigma(x)\,\sigma(x^\prime)\nonumber \\
 & & \qquad \qquad\qquad\;\; \times\,2\,g^2_{\sigma\pi\pi}\,\Big(9\,\left<0\right|T[\sigma(x)\,\sigma(x^\prime)]\left|0\right>^2+N_\pi\,\left<0\right|T[\pi(x)\,\pi(x^\prime)] \left|0\right>^2 \Big) \nonumber \\[1mm]
 & = & \frac{i}{2}\int d^4x \; \sigma(x)^2 \;\int d^4z  \;\,i \int \frac{d^4q}{(2\pi)^4} \; e^{-i q\cdot (x-z)}  \frac{1}{q^2-m^2_\sigma} \nonumber \\
 & & \quad \quad\,\;\; \times\, 6\,g_{\sigma\pi\pi}\, \Big\{   i \int \frac{d^4p}{(2\pi)^4} \;\left( -\frac{4\,g\,N_F\,m_q}{p^2-m^2_q} +\frac{3\,g_{\sigma\pi\pi}}{p^2-m_\sigma^2}+\frac{N_\pi \,g_{\sigma\pi\pi}}{p^2-m^2_\pi} \right) \nonumber \\
 & & \qquad \qquad\qquad\;\; + \,i \int d^4z^\prime \; i^3\int \frac{d^4p_1}{(2\pi)^4}\int \frac{d^4p_2}{(2\pi)^4}\int \frac{d^4p_3}{(2\pi)^4}\; e^{-i(p_1+p_2+p_3)\cdot(z-z^\prime)}\nonumber \\
 & & \qquad \qquad\;\; \times\,\left( - \frac{\lambda}{4}\right) \frac{8\,g_{\sigma\pi\pi}}{p^2_1-m^2_\sigma}\,\,\left(\frac{3}{(p^2_2-m^2_\sigma)(p^2_3-m^2_\sigma)}+\frac{N_\pi}{(p^2_2-m^2_\pi)(p^2_3-m^2_\pi)}\right)\Big\} \nonumber \\
 & - & \frac{i}{2}\int d^4x \int d^4x^\prime \;\sigma(x)\,\sigma(x^\prime)\;\,i^2\int \frac{d^4p_1}{(2\pi)^4}\int \frac{d^4p_2}{(2\pi)^4} \; e^{-i(p_1+p_2)\cdot(x-x^\prime)}\; \frac{4 \,g^2 \,N_F \,(p_1\cdot p_2 + m^2_q)}{(p_1^2-m^2_q)(p_2^2-m^2_q)} \nonumber \\[1mm]
 & + & \int d^4x \;\sigma(x)^2 \;\, 2\,\left(-\,\frac{\lambda}{4}\right)\; i\int \frac{d^4p}{(2\pi)^4}\left( \frac{3}{p^2-m^2_\sigma} + \frac{N_\pi}{p^2-m^2_\pi} \right)\nonumber \\[1mm]
 & + & \frac{i}{2}\int d^4x \; \int d^4x^\prime \; \sigma(x)\,\sigma(x^\prime) \; i^3\int \frac{d^4p_1}{(2\pi)^4}\int \frac{d^4p_2}{(2\pi)^4}\int \frac{d^4p_3}{(2\pi)^4}\; e^{-i(p_1+p_2+p_3)\cdot(x-x^\prime)} \nonumber \\
 & & \qquad \qquad\qquad\;\; \times\,\frac{2\,\lambda^2}{p^2_1-m^2_\sigma}\,\left(\frac{3}{(p^2_2-m^2_\sigma)(p^2_3-m^2_\sigma)}+\frac{N_\pi}{(p^2_2-m^2_\pi)(p^2_3-m^2_\pi)} \right) \nonumber \\[1mm]
 & + & \frac{i}{2}\int d^4x \int d^4x^\prime \; \sigma(x)\,\sigma(x^\prime) \; i^2\int \frac{d^4p_1}{(2\pi)^4}\int \frac{d^4p_2}{(2\pi)^4}\; e^{-i(p_1+p_2)\cdot(x-x^\prime)}\nonumber \\
 & & \qquad \qquad\qquad\;\; \times\,2\,g^2_{\sigma\pi\pi}\,\left(\frac{9}{(p^2_1-m^2_\sigma)(p^2_2-m^2_\sigma)}+\frac{N_\pi}{(p^2_1-m^2_\pi)(p^2_2-m^2_\pi)} \right) \nonumber \\[1mm]
 & = & \frac{i}{2}\int d^4x \; \sigma(x)^2 \;\, \frac{6\,g_{\sigma\pi\pi}}{m^2_\sigma}\, \Big\{   \int \frac{d^4p}{(2\pi)^4} \;\left( -\frac{4\,g\;N_F\,m_q}{p^2-m^2_q} +\frac{3\,g_{\sigma\pi\pi}}{p^2-m_\sigma^2}+\frac{N_\pi \,g_{\sigma\pi\pi}}{p^2-m^2_\pi} \right) \nonumber \\
 & & \qquad \qquad\qquad\qquad\quad\; + \, i\int \frac{d^4p_1}{(2\pi)^4}\int \frac{d^4p_2}{(2\pi)^4}\int \frac{d^4p_3}{(2\pi)^4}\; (2\pi)^4\, \delta^4(p_1+p_2+p_3)\nonumber \\
 & &  \qquad \qquad\qquad\qquad\quad\;\times\, \frac{2\,\lambda\, g_{\sigma\pi\pi}}{p^2_1-m^2_\sigma}\,\,\left(\frac{3}{(p^2_2-m^2_\sigma)(p^2_3-m^2_\sigma)}+\frac{N_\pi}{(p^2_2-m^2_\pi)(p^2_3-m^2_\pi)}\right)\Big\} \nonumber \\
 & + & \frac{i}{2}\int d^4x \;\sigma(x)^2\;4 \,g^2 \,N_F \int \frac{d^4p}{(2\pi)^4} \; \frac{p^2 + m^2_q}{(p^2-m^2_q)^2} \nonumber \\[1mm]
 & - & \frac{i}{2}\int d^4x \;\sigma(x)^2 \;\, \lambda \int \frac{d^4p}{(2\pi)^4}\left( \frac{3}{p^2-m^2_\sigma} + \frac{N_\pi}{p^2-m^2_\pi} \right)\nonumber \\[1mm]
 & - & \frac{i}{2}\int d^4x \; \sigma(x)^2 \; i\int \frac{d^4p_1}{(2\pi)^4}\int \frac{d^4p_2}{(2\pi)^4}\int \frac{d^4p_3}{(2\pi)^4} \; (2\pi)^4\, \delta^4(p_1+p_2+p_3) \nonumber \\
 & & \qquad \qquad\qquad\;\; \times\,\frac{2\,\lambda^2}{p^2_1-m^2_\sigma}\,\left(\frac{3}{(p^2_2-m^2_\sigma)(p^2_3-m^2_\sigma)}+\frac{N_\pi}{(p^2_2-m^2_\pi)(p^2_3-m^2_\pi)} \right) \nonumber \\[1mm]
 & - & \frac{i}{2}\int d^4x  \; \sigma(x)^2 \; \,2\,g^2_{\sigma\pi\pi}\int \frac{d^4p}{(2\pi)^4}\left(\frac{9}{(p^2-m^2_\sigma)^2}+\frac{N_\pi}{(p^2-m^2_\pi)^2} \right) \; + \; \mbox{non-local terms}\nonumber \\[1mm]
 & = & \frac{i}{2}\int d^4x \; \sigma(x)^2 \;\, \frac{6\,g_{\sigma\pi\pi}}{m^2_\sigma}\, \Big\{   - \, 4\,g \,N_F\,m_q \; I_1(m^2_q) + 3\,g_{\sigma\pi\pi}\;I_1(m^2_\sigma)+N_\pi \,g_{\sigma\pi\pi}\;I_1(m^2_\pi) \nonumber \\
 & & \qquad \qquad\qquad\qquad \;\;\;\;\, +\, 2\,\lambda\,g_{\sigma\pi\pi} \; i\,\Big(3\;I^{\,sunset}_{1,1,1}(m^2_\sigma,m^2_\sigma,m^2_\sigma)+N_\pi\;I^{\,sunset}_{1,1,1}(m^2_\sigma,m^2_\pi,m^2_\pi)\Big)\Big\} \nonumber \\
 & + & \frac{i}{2}\int d^4x \;\sigma(x)^2\;4 \,g^2 \,N_F  \; \Big( I_1(m^2_q) + 2\, m^2_q \;I_2(m^2_q) \Big) \nonumber \\[1mm]
 & - & \frac{i}{2}\int d^4x \;\sigma(x)^2 \;\, \lambda \;\Big( 3\; I_1(m^2_\sigma) + N_\pi\; I_1(m^2_\pi) \Big)\nonumber \\[1mm]
 & - & \frac{i}{2}\int d^4x \; \sigma(x)^2 \; 2\,\lambda^2\; i\;\Big(3\;I^{\,sunset}_{1,1,1}(m^2_\sigma,m^2_\sigma,m^2_\sigma)+N_\pi\;I^{\,sunset}_{1,1,1}(m^2_\sigma,m^2_\pi,m^2_\pi) \Big) \nonumber \\[1mm]
 & - & \frac{i}{2}\int d^4x  \; \sigma(x)^2 \; \,2\,g^2_{\sigma\pi\pi}\;\Big(9\;I_2(m^2_\sigma)+N_\pi\;I_2(m^2_\pi) \Big)\quad + \quad  \mbox{non-local terms}\; . \label{eqeffact3}
\end{eqnarray}
Analogously the effective action $S_{(4)}[\vec{\pi}^2]$ for the two-point function of the $\pi$ consisting of the contributions illustrated in Fig.\ \ref{fig4} is given by  \cite{Kleefeld:2005hd}:
\begin{eqnarray} \lefteqn{S_{(4)}[\vec{\pi}^2]=} \nonumber \\[2mm]
 & + & S_{(4a)}[\vec{\pi}^2] +S_{(4b)}[\vec{\pi}^2] +S_{(4c)}[\vec{\pi}^2] +S_{(4d)}[\vec{\pi}^2] +S_{(4e)}[\vec{\pi}^2]\nonumber \\
 & + & S_{(4f)}[\vec{\pi}^2] +S_{(4g)}[\vec{\pi}^2] +S_{(4h)}[\vec{\pi}^2] +S_{(4i)}[\vec{\pi}^2] +S_{(4j)}[\vec{\pi}^2]+S_{(4k)}[\vec{\pi}^2] +S_{(4l)}[\vec{\pi}^2] \nonumber \\[1mm]
 & = & \frac{i}{2}\int d^4x \; \vec{\pi}(x)^2 \;\int d^4z \;   \left<0\right|T[\sigma(x)\,\sigma(z)]\left|0\right>\nonumber \\
 & & \quad \quad\,\;\; \times\, 2\,g_{\sigma\pi\pi}\, \Big\{  \left<0\right|T\Big[ g \; \overline{q^c_+}(z)\,q_-(z)+ 3\,g_{\sigma\pi\pi}\,\sigma(z)^2+N_\pi \,g_{\sigma\pi\pi}\,\pi(z)^2 \Big]\left|0\right> \nonumber \\
 & & \qquad \qquad\qquad\;\; + \,i \int d^4z^\prime \, \left( - \frac{\lambda}{4}\right)\,g_{\sigma\pi\pi}\left<0\right|T[\sigma(z)\,\sigma(z^\prime)]\left|0\right>\nonumber \\
 & & \qquad \qquad\qquad\;\; \times\,8\,\Big(3\,\left<0\right|T[\sigma(z)\,\sigma(z^\prime)]\left|0\right>^2+N_\pi\,\left<0\right|T[\pi(z)\,\pi(z^\prime)] \left|0\right>^2 \Big)\Big\} \nonumber \\[1mm]
 & + & \frac{i}{2}\int d^4x \int d^4x^\prime \;\vec{\pi}(x)\cdot \vec{\pi}(x^\prime)\,g^2 \left<0\right|T\Big[ \overline{q^c_+}(x)\;i\,\gamma_5\;q_-(x)\;\overline{q^c_+}(x^\prime)\;i\,\gamma_5\;q_-(x^\prime) \Big]\left|0\right>_c  \nonumber \\[1mm]
 & + & \int d^4x \;\vec{\pi}(x)^2 \;\, 2\,\left(-\,\frac{\lambda}{4}\right)\, \Big( \left<0\right|T[\sigma(x)\,\sigma(x)]\left|0\right> + (N_\pi+2) \,\left<0\right|T[\pi(x)\,\pi(x)]\left|0\right>\Big)\nonumber \\[1mm]
 & + & \frac{i}{2}\int d^4x \; \int d^4x^\prime \; \vec{\pi}(x)\cdot\vec{\pi}(x^\prime) \;  \left<0\right|T[\pi(x)\,\pi(x^\prime)]\left|0\right>\nonumber \\
 & & \qquad \qquad\qquad\;\; \times\,2\,\lambda^2\,\Big(\left<0\right|T[\sigma(x)\,\sigma(x^\prime)]\left|0\right>^2+(N_\pi+2)\,\left<0\right|T[\pi(x)\,\pi(x^\prime)] \left|0\right>^2 \Big) \nonumber \\[1mm]
 & + & \frac{i}{2}\int d^4x \int d^4x^\prime \; \vec{\pi}(x)\cdot\vec{\pi}(x^\prime)\;\, 4\,g^2_{\sigma\pi\pi}\,\left<0\right|T[\sigma(x)\,\sigma(x^\prime)]\left|0\right> \,\left<0\right|T[\pi(x)\,\pi(x^\prime)] \left|0\right> \nonumber \\[1mm]
 & = & \frac{i}{2}\int d^4x \; \vec{\pi}(x)^2 \;\int d^4z  \;\,i \int \frac{d^4q}{(2\pi)^4} \; e^{-i q\cdot (x-z)}  \frac{1}{q^2-m^2_\sigma} \nonumber \\
 & & \quad \quad\,\;\; \times\, 2\,g_{\sigma\pi\pi}\, \Big\{   i \int \frac{d^4p}{(2\pi)^4} \;\left( -\frac{4\,g\,N_F\,m_q}{p^2-m^2_q} +\frac{3\,g_{\sigma\pi\pi}}{p^2-m_\sigma^2}+\frac{N_\pi \,g_{\sigma\pi\pi}}{p^2-m^2_\pi} \right) \nonumber \\
 & & \qquad \qquad\qquad\;\; + \,i \int d^4z^\prime \; i^3\int \frac{d^4p_1}{(2\pi)^4}\int \frac{d^4p_2}{(2\pi)^4}\int \frac{d^4p_3}{(2\pi)^4}\; e^{-i(p_1+p_2+p_3)\cdot(z-z^\prime)}\nonumber \\
 & & \qquad \qquad\;\; \times\,\left( - \frac{\lambda}{4}\right) \frac{8\,g_{\sigma\pi\pi}}{p^2_1-m^2_\sigma}\,\,\left(\frac{3}{(p^2_2-m^2_\sigma)(p^2_3-m^2_\sigma)}+\frac{N_\pi}{(p^2_2-m^2_\pi)(p^2_3-m^2_\pi)}\right)\Big\} \nonumber \\
 & - & \frac{i}{2}\int d^4x \int d^4x^\prime \;\vec{\pi}(x)\cdot\vec{\pi}(x^\prime)\;\,i^2\int \frac{d^4p_1}{(2\pi)^4}\int \frac{d^4p_2}{(2\pi)^4} \; e^{-i(p_1+p_2)\cdot(x-x^\prime)}\; \frac{4 \,g^2 \,N_F \,(p_1\cdot p_2 - m^2_q)}{(p_1^2-m^2_q)(p_2^2-m^2_q)} \nonumber \\[1mm]
 & + & \int d^4x \;\vec{\pi}(x)^2 \;\, 2\,\left(-\,\frac{\lambda}{4}\right)\; i\int \frac{d^4p}{(2\pi)^4}\left( \frac{1}{p^2-m^2_\sigma} + \frac{N_\pi+2}{p^2-m^2_\pi} \right)\nonumber \\[1mm]
 & + & \frac{i}{2}\int d^4x \; \int d^4x^\prime \; \vec{\pi}(x)\cdot\vec{\pi}(x^\prime) \; i^3\int \frac{d^4p_1}{(2\pi)^4}\int \frac{d^4p_2}{(2\pi)^4}\int \frac{d^4p_3}{(2\pi)^4}\; e^{-i(p_1+p_2+p_3)\cdot(x-x^\prime)} \nonumber \\
 & & \qquad \qquad\qquad\;\; \times\,\frac{2\,\lambda^2}{p^2_1-m^2_\pi}\,\left(\frac{1}{(p^2_2-m^2_\sigma)(p^2_3-m^2_\sigma)}+\frac{N_\pi+2}{(p^2_2-m^2_\pi)(p^2_3-m^2_\pi)} \right) \nonumber \\[1mm]
 & + & \frac{i}{2}\int d^4x \int d^4x^\prime \; \vec{\pi}(x)\cdot\vec{\pi}(x^\prime) \; i^2\int \frac{d^4p_1}{(2\pi)^4}\int \frac{d^4p_2}{(2\pi)^4}\; e^{-i(p_1+p_2)\cdot(x-x^\prime)}\nonumber \\
 & & \qquad \qquad\qquad\;\; \times\,4\,g^2_{\sigma\pi\pi}\;\frac{1}{(p^2_1-m^2_\sigma)(p^2_2-m^2_\pi)} \nonumber \\[1mm]
 & = & \frac{i}{2}\int d^4x \; \vec{\pi}(x)^2 \;\, \frac{2\,g_{\sigma\pi\pi}}{m^2_\sigma}\, \Big\{   \int \frac{d^4p}{(2\pi)^4} \;\left( -\frac{4\,g\;N_F\,m_q}{p^2-m^2_q} +\frac{3\,g_{\sigma\pi\pi}}{p^2-m_\sigma^2}+\frac{N_\pi \,g_{\sigma\pi\pi}}{p^2-m^2_\pi} \right) \nonumber \\
 & & \qquad \qquad\qquad\qquad\quad\; + \, i\int \frac{d^4p_1}{(2\pi)^4}\int \frac{d^4p_2}{(2\pi)^4}\int \frac{d^4p_3}{(2\pi)^4}\; (2\pi)^4\, \delta^4(p_1+p_2+p_3)\nonumber \\
 & &  \qquad \qquad\qquad\qquad\quad\;\times\, \frac{2\,\lambda\, g_{\sigma\pi\pi}}{p^2_1-m^2_\sigma}\,\,\left(\frac{3}{(p^2_2-m^2_\sigma)(p^2_3-m^2_\sigma)}+\frac{N_\pi}{(p^2_2-m^2_\pi)(p^2_3-m^2_\pi)}\right)\Big\} \nonumber \\
 & + & \frac{i}{2}\int d^4x \;\vec{\pi}(x)^2\;\,4 \,g^2 \,N_F \,\int \frac{d^4p}{(2\pi)^4} \; \frac{1}{p^2-m^2_q} \nonumber \\[1mm]
 & - & \frac{i}{2}\int d^4x \;\vec{\pi}(x)^2 \;\, \lambda \int \frac{d^4p}{(2\pi)^4}\left( \frac{1}{p^2-m^2_\sigma} + \frac{N_\pi+2}{p^2-m^2_\pi} \right)\nonumber \\[1mm]
 & - & \frac{i}{2}\int d^4x \; \vec{\pi}(x)^2 \; i\int \frac{d^4p_1}{(2\pi)^4}\int \frac{d^4p_2}{(2\pi)^4}\int \frac{d^4p_3}{(2\pi)^4} \; (2\pi)^4\, \delta^4(p_1+p_2+p_3) \nonumber \\
 & & \qquad \qquad\qquad\;\; \times\,\frac{2\,\lambda^2}{p^2_1-m^2_\pi}\,\left(\frac{1}{(p^2_2-m^2_\sigma)(p^2_3-m^2_\sigma)}+\frac{N_\pi+2}{(p^2_2-m^2_\pi)(p^2_3-m^2_\pi)} \right) \nonumber \\[1mm]
 & - & \frac{i}{2}\int d^4x  \; \vec{\pi}(x)^2 \; \,4\,g^2_{\sigma\pi\pi}\int \frac{d^4p}{(2\pi)^4}\;\frac{1}{(p^2-m^2_\sigma)(p^2-m^2_\pi)} \; + \;  \mbox{non-local terms} \nonumber \\[3mm]
 & = & \frac{i}{2}\int d^4x \; \vec{\pi}(x)^2 \;\, \frac{2\,g_{\sigma\pi\pi}}{m^2_\sigma}\, \Big\{  - \, 4\,g \,N_F\,m_q \; I_1(m^2_q) + 3\,g_{\sigma\pi\pi}\;I_1(m^2_\sigma)+N_\pi \,g_{\sigma\pi\pi}\;I_1(m^2_\pi) \nonumber \\
 & & \qquad \qquad\qquad\qquad \;\;\;\;\, +\, 2\,\lambda\,g_{\sigma\pi\pi} \; i\,\Big(3\;I^{\,sunset}_{1,1,1}(m^2_\sigma,m^2_\sigma,m^2_\sigma)+N_\pi\;I^{\,sunset}_{1,1,1}(m^2_\sigma,m^2_\pi,m^2_\pi)\Big) \Big\} \nonumber \\
 & + & \frac{i}{2}\int d^4x \;\vec{\pi}(x)^2\;\,4 \,g^2 \,N_F  \; I_1(m^2_q) \nonumber \\[1mm]
 & - & \frac{i}{2}\int d^4x \;\vec{\pi}(x)^2 \;\, \lambda \; \Big( I_1(m^2_\sigma) + (N_\pi+2)\; I_1(m^2_\pi) \Big)\nonumber \\[1mm]
 & - & \frac{i}{2}\int d^4x \; \vec{\pi}(x)^2 \; 2\,\lambda^2\; i\;\Big(I^{\,sunset}_{1,1,1}(m^2_\pi,m^2_\sigma,m^2_\sigma)+(N_\pi+2)\;I^{\,sunset}_{1,1,1}(m^2_\pi,m^2_\pi,m^2_\pi)\Big)\nonumber \\[1mm]
 & - & \frac{i}{2}\int d^4x  \; \vec{\pi}(x)^2 \; \,4\,g^2_{\sigma\pi\pi} \; I_{1,1}(m^2_\sigma,m^2_\pi) \; + \;  \mbox{non-local terms}\; . \label{eqeffact4}
\end{eqnarray}
For convenience we want to recall here also the derivation of the quark-loop contribution to the effective actions for the $\sigma\pi\pi$- and the $\pi^4$-interactions  \cite{Kleefeld:2005hd}:
\begin{eqnarray} \lefteqn{S_{\mbox{\small quark-loop}}[\sigma \,\vec{\pi}^2]=\frac{i^2}{2!} \int d^4x \int d^4x_1\int d^4x_2 \;(-2)\; g^3 \;\mbox{tr}\Big[ \sigma(x)\, \left<0\right|T[ q_-(x)\;\overline{q^c_+}(x_1) ]\left|0\right>_c} \nonumber \\[2mm]
 &  & \times \;i\,\gamma_5\;\vec{\tau}\cdot \vec{\pi}(x_1)\; \left<0\right|T[ q_-(x_1)\;\overline{q^c_+}(x_2)]\left|0\right>_c\;i\,\gamma_5\;\vec{\tau}\cdot \vec{\pi}(x_2)\; \left<0\right|T[ q_-(x_2) \; \overline{q^c_+}(x)]\left|0\right>_c \Big] \nonumber \\[1mm]
 & = & \frac{i^2}{2!} \int d^4x \int d^4x_1\int d^4x_2 \;\,  (-2)\;  g^3 \, N_F \; \sigma(x)\;  \vec{\pi}(x_1)\cdot \vec{\pi}(x_2) \nonumber \\
 & & \times\; i^3 \int \frac{d^4p_{{}_{01}}}{(2\pi)^4}\int \frac{d^4p_{{}_{12}}}{(2\pi)^4}\int \frac{d^4p_{{}_{20}}}{(2\pi)^4} \;\; e^{-i\,p_{{}_{01}}\cdot (x-x_1)}\; e^{-i\,p_{{}_{12}}\cdot (x_1-x_2)}\; e^{-i\,p_{{}_{20}}\cdot (x_2-x)}\nonumber \\[2mm]
 &  & \times \; \mbox{tr}\left[ \frac{\not\! p_{{}_{01}} + m_q}{p^2_{{}_{01}}-m^2_q} \;i\,\gamma_5\; \frac{\not\! p_{{}_{12}} + m_q}{p^2_{{}_{12}}-m^2_q}\;i\,\gamma_5\;\frac{\not\! p_{{}_{20}} + m_q}{p^2_{{}_{20}}-m^2_q}  \right] \nonumber \\[1mm]
 & = & i \int d^4x \;\,  (-1)\, g^3 \, N_F \; \sigma(x)\;  \vec{\pi}(x)^2 \;  \int \frac{d^4p}{(2\pi)^4} \;\; \mbox{tr}\left[ \frac{\not\! p + m_q}{p^2-m^2_q} \;i\,\gamma_5\; \frac{\not\! p + m_q}{p^2-m^2_q}\;i\,\gamma_5\;\frac{\not\! p + m_q}{p^2-m^2_q}  \right] \nonumber \\[1mm]
 & + & \mbox{non-local terms} \nonumber \\[1mm]
 & = & i \int d^4x \;\,  (-4)\, g^3 \, N_F\,m_q \; \sigma(x)\;  \vec{\pi}(x)^2 \;  \int \frac{d^4p}{(2\pi)^4} \;\;  \frac{1}{(p^2-m^2_q)^2} \; + \; \mbox{non-local terms}\nonumber \\[1mm]
 & = & i \int d^4x \;\,  (-4)\, g^3 \, N_F\,m_q \; \sigma(x)\;  \vec{\pi}(x)^2 \;  I_2(m^2_q) \; + \; \mbox{non-local terms}\; , \label{eqqloop1} \\[1mm]
\lefteqn{S_{\mbox{\small quark-loop}}[(\vec{\pi}^2)^2]= \frac{i^3}{4!} \int d^4x \int d^4x_1\int d^4x_2 \int d^4x_3 \;(-6)\; g^4 } \nonumber \\[2mm]
 & & \times \; \mbox{tr}\Big[ i\,\gamma_5\;\vec{\tau}\cdot \vec{\pi}(x)\; \left<0\right|T[ q_-(x)\;\overline{q^c_+}(x_1) ]\left|0\right>_c\; i\,\gamma_5\;\vec{\tau}\cdot \vec{\pi}(x_1)\; \left<0\right|T[ q_-(x_1)\;\overline{q^c_+}(x_2) ]\left|0\right>_c\nonumber \\[2mm]
 &  & \quad \times \;i\,\gamma_5\;\vec{\tau}\cdot \vec{\pi}(x_2)\;  \left<0\right|T[ q_-(x_2)\;\overline{q^c_+}(x_3)]\left|0\right>_c\;i\,\gamma_5\;\vec{\tau}\cdot \vec{\pi}(x_3)\;  \left<0\right|T[ q_-(x_3) \; \overline{q^c_+}(x)]\left|0\right>_c \Big] \nonumber \\[1mm]
 & = & \frac{i^3}{4!} \int d^4x \int d^4x_1\int d^4x_2 \;\,  (-6)\;  g^4 \, N_F \;  \vec{\pi}(x)\cdot \vec{\pi}(x_1)\;  \vec{\pi}(x_2)\cdot \vec{\pi}(x_3) \nonumber \\
 & & \times\; i^4 \int \frac{d^4p_{{}_{01}}}{(2\pi)^4}\int \frac{d^4p_{{}_{12}}}{(2\pi)^4}\int \frac{d^4p_{{}_{23}}}{(2\pi)^4}\int \frac{d^4p_{{}_{30}}}{(2\pi)^4} \; \; e^{-i\,p_{{}_{01}}\cdot (x-x_1)}\; e^{-i\,p_{{}_{12}}\cdot (x_1-x_2)}\; e^{-i\,p_{{}_{23}}\cdot (x_2-x_3)}\nonumber \\[2mm]
 & & \times\;  \; e^{-i\,p_{{}_{30}}\cdot (x_3-x)} \; \mbox{tr}\left[i\,\gamma_5\; \frac{\not\! p_{{}_{01}} + m_q}{p^2_{{}_{01}}-m^2_q} \;i\,\gamma_5\; \frac{\not\! p_{{}_{12}} + m_q}{p^2_{{}_{12}}-m^2_q}\;i\,\gamma_5\;\frac{\not\! p_{{}_{23}} + m_q}{p^2_{{}_{23}}-m^2_q}\;i\,\gamma_5\;\frac{\not\! p_{{}_{30}} + m_q}{p^2_{{}_{30}}-m^2_q}  \right] \nonumber \\[1mm]
 & = & \frac{i}{4} \int d^4x \;\, g^4 \, N_F \;  \Big( \vec{\pi}(x)^2\Big)^2 \;  \int \frac{d^4p}{(2\pi)^4}  \nonumber \\[1mm]
 & & \times  \; \mbox{tr}\left[ i\,\gamma_5\;\frac{\not\! p + m_q}{p^2-m^2_q} \;i\,\gamma_5\; \frac{\not\! p + m_q}{p^2-m^2_q}\;i\,\gamma_5\;\frac{\not\! p + m_q}{p^2-m^2_q} \;i\,\gamma_5\;\frac{\not\! p + m_q}{p^2-m^2_q} \right] \;+ \; \mbox{non-local terms} \nonumber \\[1mm]
 & = & i \int d^4x \;\,   g^4 \, N_F \; \Big( \vec{\pi}(x)^2\Big)^2  \;  \int \frac{d^4p}{(2\pi)^4} \;\;  \frac{1}{(p^2-m^2_q)^2} \; + \; \mbox{non-local terms} \nonumber \\[1mm]
 & = & i \int d^4x \;\,   g^4 \, N_F \; \Big( \vec{\pi}(x)^2\Big)^2  \;  I_2(m^2_q) \; + \; \mbox{non-local terms}\; .\label{eqqloop2}
\end{eqnarray}
\end{appendix}

\end{document}